\documentclass[12 pt]{article}        
\usepackage{epsfig}
\usepackage{cite}
\usepackage{graphics}
\usepackage{color}

\renewcommand{\thefootnote}{\fnsymbol{footnote}}

\begin{document}                     


\begin{titlepage}

\begin{flushright}
 {\bf TPJU 2/2001}
\end{flushright}
\vspace{1mm}
\begin{center}

\vspace{1cm}
{\LARGE \bf  Spin effects in $\tau$ lepton pair production}\\
\vspace{0.5cm}
{\LARGE \bf  at LHC~\footnote[2]{~ Work supported in part by
  the Polish State Committee grants KBN 2P03B11819  2P03B05418 \\and
by the European Commission 5-th Framework contract HPRN-CT-2000-00149. }}
\end{center}
\vspace{1cm}

\begin{center}
{\large \bf T. Pierzcha\l a$^{a}$ , E. Richter-W\c as$
^{b,c,d}$, Z.W\c as$^{d,e}$
 and M. Worek$^{a}$}\

\vspace{5mm}
$^a${\em Institute of  Physics, University of Silesia \\
 Uniwersytecka 4, 40-007 Katowice, Poland}\\
$^b$ {\em Institute of Computer Science, Jagellonian University \\
Nawojki 11, 30-072 Cracow, Poland}\\
$^c${\em CERN, PPE, 1211 Geneva 23, Switzerland}\\

$^d${\em Institute of Nuclear Physics\\
   Kawiory 26a, 30-055 Cracow, Poland}\\
$^e${\em CERN, Theory Division, 1211 Geneva 23, Switzerland}
\end{center}

\vspace{1cm}
\begin{abstract}
The proper incorporation of 
spin effects in $\tau$ lepton decays is often of importance. 
In present work the case of the
$Z/\gamma \to  \tau^+ \tau^-$ production mechanism is studied 
in detail. As an example, the effects due to the spin correlations  
on the  potential for the
Minimal Supersymmetric Standard Model (MSSM)
Higgs boson(s) searches in the $\tau \tau$ decay 
channel at  the Large Hadron Collider (LHC) are discussed. 
For these processes, the Standard Model 
$Z/\gamma^* \to \tau \tau$-pair production 
is a dominant background. The spin effects in high energy 
physics reactions, can be implemented 
up to certain approximation, independently
of the algorithm and matrix elements used by the production program. 
Information stored on every generated event can be sufficient. 
The algorithm based on such approximation is documented.
Question of the theoretical uncertaintity is partly discussed.
\end{abstract}

 \vspace{1cm}
\centerline{ {\it Acta Physica Polonica} {\bf B 32} (2001) 1277}

\vfill
\begin{flushleft}
{ \bf  January 2001}
\end{flushleft}

\end{titlepage}

\renewcommand{\thefootnote}{\arabic{footnote}}
\section{\bf \large   Introduction}
\vskip 0.3 cm
In a study of ``discovery potential'' and data analysis of present high 
energy
experiments the problems of precise predictions including, simultaneously,
signal signatures of the new (or studied) physics, backgrounds, as well
as all detector related effects should be analysed. 
It is generally believed that a Monte Carlo simulation 
of the full chain from the beam collision to detector response is the most 
convenient technique to address such question.
In general it is indispensable to divide Monte Carlo simulation into
separate blocks: physics event generation and detector response.
Later event generation can be divided further into parts, 
describing for example production and decay of the intermediate states.

In the present paper we will concentrate on the particular class 
of the processes involving polarised  $\tau$ leptons. 
The two main goals of the present paper are:
$(i)$ presentation of the algorithm for  matching 
$\tau$ lepton decay and its production, with some control
over spin effects; in particular  in case of $Z/\gamma \to \tau^+ \tau^-$
 production mechanism, $(ii)$ discussion of physical observables
 sensitive to the spin correlations in the $\tau$ pair production.

 Spin correlations
in the decay of  $\tau$ leptons not only can help to suppress
irreducible background to the possible resonant $\tau$ pair production at LHC,
such as the MSSM Higgs bosons decays, but also help to determine
the spin nature of this resonanse.

In the papers ~\cite{tauola:1990,tauola:1992,tauola:1993} 
{\tt TAUOLA} Monte Carlo package for
simulation of $\tau$ lepton decay was described. 
Recently, in Ref.~\cite{Golonka:2000iu}, technical details 
convenient for using the code in multi-purpose environment were collected,
and universal interface for combining the simulation of  $\tau$
lepton decay, with different packages for generation of physics event 
was proposed. Scheme of Ref.~\cite{Golonka:2000iu} relies on the
information stored in the {\tt HEPEVT} common block \cite{PDG:1998} only,
and not on the details specific for the production generator, such as 
{\tt PYTHIA} \cite{Pythia} (used in our examples). In fact, such an 
interface can be considered as a separate software project, to some degree
 independent both from
the specific problem of $\tau$ production and its decay.
  
Our paper is organized as follows:
in the next section
we will describe  new algorithm for extracting elementary 
$2 \to 2 $ body reaction for  $f \bar f  \to Z/\gamma \to \tau^+ \tau^-$, which
is necessary for properly introducing spin correlations into generation chain.
In Sec. 3 we analyze spin content of such an elementary function. Sec. 4 
is dedicated to the discussion of  their consequences for the distributions 
of physics interest.
In Sec. 5 we discuss few observables where spin effects can improve
 separation of 
the Higgs boson signature, in case of the 14 TeV $pp$ collisions.
Summary closes the paper.
In Appendix, we explain the basic scheme of the spin treatment used in our
 code. It completes the program manual given in Ref.~\cite{Golonka:2000iu}.

\section{The $\tau$ polarisation from the $Z/\gamma \to \tau^+ \tau^- $ decay}

The exact way of calculating spin state of any final state is 
with the help of the matrix element and the rigorous density matrix treatment. 
This is however not always possible or necessary. Often, like in the
case of the production and decay of particles in the ultra-relativistic limit a 
simplified approach can be sufficient. Such an approach was developed 
for {\tt KORALZ} Monte Carlo program \cite{Jadach:1994yv} and
its limitations were studied
with the help of  matrix element calculations of the order $\alpha$ 
\cite{koralzearly}.
In the following, we study the question whwter the approach can 
be generalised, and the approximate spin correlation calculated from the 
information stored in the {\tt HEPEVT} common block filled by ``any'' $\tau$
production program.

The approximation consists of reconstructing
information of the elementary $2 \to 2$ body 
process $e^+ e^- (q\bar q) \to \tau^+\tau^- $, buried inside multi-body
production process. Let us stress that such a procedure can never be fully
controlled, as its  functioning depends on the way the
production program fills the {\tt HEPEVT} common block. It will be 
always  responsibility of the user to check if in the particular case 
the implemented algorithm is applicable. Nonetheless our aim is {\it not} to replace
the matrix element calculations, but rather to provide a  method 
of calculating/estimating spin effects in cases when spin effects 
would not be taken care of, at all.
Needless to say  such an approach is limited (for the spin treatment) to the
approximation not better than leading-log, and to the longitudinal spin degrees only.

The principle of calculating kinematic variables is simple. 
The 4-momenta of the  $2 \to 2$ body process have to be found.
The 4-momenta of the outcoming $\tau$'s are used directly. 
Initial state momenta are constructed from the incoming and outcoming 
momenta of the particles (or fields) accompanying production of the 
$Z/\gamma$ state\footnote{The $Z/\gamma$ state  does not need 
to be explicitly coded in the {\tt HEPEVT} common block. Note that if 
available, information from the history part of the event, where the 4-momenta
of gluons quarks etc. are stored, will be used.}. 
We group them accordingly to fermion number flow, and ambiguous additional 
particles are grouped (summed) into effective quarks  to minimise their 
virtualities. Such an approach is internally consistent in the case of
 emission of photons or gluons within the leading log approximation.

Longitudinal polarisation of $\tau$ leptons 
 ${\cal P}_{\tau}$ depends on the spin quantum number
of the $\tau$ mother \footnote{The spin quantisation axes are chosen {\it in the same} way as in Ref.~\cite{Jadach:1994yv}.}. It is randomly generated 
as specified in Table~\ref{T:Probability}.

\begin{table}
\newcommand{\lstrut}{{$\strut\atop\strut$}}
  \caption {\em Probability for the configurations of the longitudinal polarisation of the pair of  $\tau$ leptons from different origins. 
\label{T:Probability}}
\vspace{2mm}
\begin{center}
\begin{tabular}{|c|c|c|c|} \hline \hline 
Origin & $P_{\tau^{+}}$  & $ P_{\tau^{-}}$ & Probability \\	\hline
 Neutral Higgs bosons: $h^{0},H^{0},A^{0}$ &   
 $P_{\tau^{+}}=+1$ & $P_{\tau^{-}}=-1$  & 0.5 \\
& $P_{\tau^{+}}=-1$ & $P_{\tau^{-}}=+1$  & 0.5 \\ \hline
 Charged Higgs boson: $ H^{+}$ or $H^{-}$ &
 $P_{\tau^{+}}=+1$ & $P_{\tau^{-}}=+1$  & 1.0 \\ \hline
 Charged vector boson:  $W^{+}$ or $W^{-}$ &
 $P_{\tau^{+}}=-1$ & $P_{\tau^{-}}=-1$  & 1.0  \\ \hline
 Neutral vector boson: $Z/\gamma^{*}$ &
 $P_{\tau^{+}}=+1$ & $P_{\tau^{-}}=+1$  & $P_{Z}$ \\
& $P_{\tau^{+}}=-1$ & $P_{\tau^{-}}=-1$  & $1-P_{Z}$  \\ \hline 
 Other  &
 $P_{\tau^{+}}=+1$ & $P_{\tau^{-}}=+1$  & 0.5 \\
& $P_{\tau^{+}}=-1$ & $P_{\tau^{-}}=-1$  & 0.5 \\ \hline \hline
\end{tabular}
\end{center}
\end{table}
\newpage
The probability $P_{Z}$ used in the generation, is calculated directly from the squares 
of the matrix elements of the Born-level 
$2 \to 2 $ process
$f \bar f  \to \tau^{- }{\tau}^{+ }$:
\begin{equation}
P_{Z}=\frac{\left|{\cal M}\right|^2_{f \bar{f} \to \tau^{-}
\tau^{+ }}\left(+ ,+\right)}
{\left|{\cal M}\right|^2_{f \bar{f} \to \tau^{-} \tau^{+ } }
\left(+ ,+\right)\ +\ 
\left|{\cal M}\right|^2_{f \bar{f} \to \tau^{-} \tau^{+ }  }
\left(- ,-\right)}\; ,
\end{equation}
where $f=e,\mu,u,d,c,s,b$. It can be also expressed
(following conventions of Ref.~\cite{Eberhard:1989ve}), 
with the help of the vector (and axial) couplings of fermions to the 
$\gamma $ ( and $Z$) bosons. Explicit expression 
for  the differential distributions is used.
The resulting formula, given below,  will be useful 
for the discussion of the numerical 
results and tests of the next sections.
\def\Born{{\rm Born}}
\begin{eqnarray}
P_{Z}(s,\theta )&=& {
{d\sigma_\Born \over d\cos\theta }
     \big(s,\cos\theta;1 \big)
\over
{d\sigma_\Born \over d\cos\theta }
     \big(s,\cos\theta;1 \big) +
{d\sigma_\Born \over d\cos\theta }
     \big(s,\cos\theta;-1 \big)
}
\\
{d\sigma_\Born \over d\cos\theta }
     \big(s,\cos\theta;p \big) &=& 
      (1+\cos^2\theta) F_0(s) +2\cos\theta F_1(s) \nonumber \\
&&-p\big[(1+\cos^2\theta) F_2(s) +2\cos\theta F_3(s)\big].
\end{eqnarray}

with the four form-factors
\def\qel{q_f} \def\qta{q_\tau}
\def\vel{v_f} \def\vta{v_\tau}
\def\ael{a_f} \def\ata{a_\tau}

\begin{eqnarray}
F_0(s)&=& {\pi\alpha^2\over 2s}
 \big( \qel^2\qta^2
       + 2\hbox{Re}\chi(s)   \qel\qta\vel\vta
       + \vert\chi(s)\vert^2 (\vel^2+\ael^2)(\vta^2+\ata^2)
  \big),
\nonumber \\
F_1(s)&=& {\pi\alpha^2\over 2s}
 \big( \qquad
         2\hbox{Re}\chi(s)   \qel\qta\ael\ata
       + \vert\chi(s)\vert^2 \; 2\vel\ael \; 2\vta\ata
  \big),
\\
F_2(s)&=& {\pi\alpha^2\over 2s}
 \big( \qquad
         2\hbox{Re}\chi(s)   \qel\qta\vel\ata
       + \vert\chi(s)\vert^2 (\vel^2+\ael^2)\; 2\vta\ata
  \big),
\nonumber \\
F_3(s)&=& {\pi\alpha^2\over 2s}
 \big( \qquad
         2\hbox{Re}\chi(s)   \qel\qta\ael\vta
       + \vert\chi(s)\vert^2 \; 2\vel\ael \; (\vta^2+\ata^2)
  \big),\nonumber 
\end{eqnarray}
and
\begin{equation}
\chi(s)= {s\over s-M_Z^2 +is\Gamma_Z/M_Z } \nonumber  .
\end{equation}
Analytic expression for the coupling constants, and  their numerical values
are given in Table \ref{Z:couplings}.
\begin{table}
\newcommand{\lstrut}{{$\strut\atop\strut$}}
  \caption {\em The $\gamma, Z$ couplings to fermions. Lowest order approximation, the numerical values are given for the effective $\sin^2\theta_W=0.23147$. 
\label{Z:couplings}}
\vspace{2mm}
\begin{center}
\begin{tabular}{|c|c|c|c|} \hline \hline 
 Flavour: $f$              & $q_f$   & $ v_{f}$ & $ a_{f}$  \\  \hline
 Leptons: $f=e, \mu, \tau$ & 1       & ${-1+4\sin^2\theta_W \over 4\sin\theta_W \cos\theta_W}=-0.044$      &  $-{1 \over 4\sin\theta_W \cos\theta_W}=-0.593$  \\ 
 \hline
 Quarks : $f= u$ or $c$    & 2/3     & ${1-{8 \over 3}\sin^2\theta_W \over 4\sin\theta_W \cos\theta_W}=0.227$      & $ {1 \over 4\sin\theta_W \cos\theta_W}=0.593$      \\ \hline
 Quarks : $f= d, s$ or $b$ & -1/3    & ${ -1+{4\over 3}\sin^2\theta_W\over 4\sin\theta_W \cos\theta_W}=-0.410$      & $-{1 \over 4\sin\theta_W \cos\theta_W}=-0.593$     \\  \hline  \hline
\end{tabular}
\end{center}
\end{table}

In the first step of our discussion the   
${\cal P}_{\tau}$ is shown as a function of $\cos{\theta}$, for several
centre of mass energies and initial state flavours.
 The angle
${\theta}$ denotes  $\tau^-$ scattering angle in the $Z/\gamma^{*}$
rest-frame. It is calculated with respect to the $e^-$, $u$ or $d$
effective beam. 
The $Z$ mass $m_Z$~=~91.1882 GeV, was taken from Ref.~\cite{PDG2000}, 
as well as
effective $\sin^{2}\theta_W=0.23147$ and $\Gamma_Z$~=~2.49~GeV.

\begin{figure}[!ht]
\centering
\setlength{\unitlength}{0.1mm}
\begin{picture}(1400,1400)
\put( 800,1300){\makebox(0,0)[b]{ $e^-e^+ \to \tau^+ \tau^-$}}
\put( 450, 600){\makebox(0,0)[b]{ $u \bar{u}\to \tau^+ \tau^- $ }}
\put(1200, 600){\makebox(0,0)[b]{  $d \bar{d}\to \tau^+ \tau^-$ }}
\put(330, 480){\makebox(0,0)[lb]{\epsfig{file=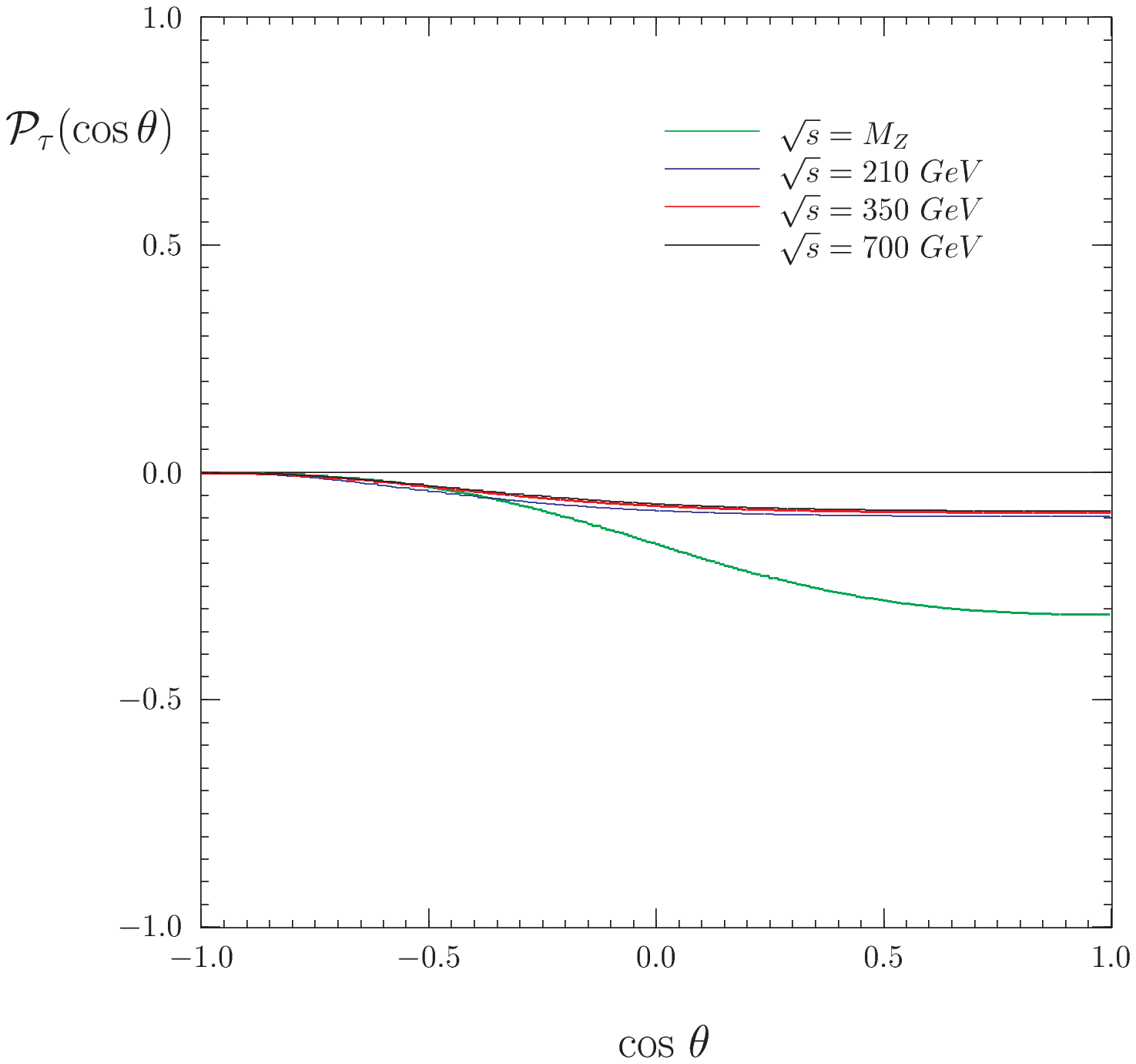,width=80mm,height=100mm}}}
\put(-20, -220){\makebox(0,0)[lb]{\epsfig{file=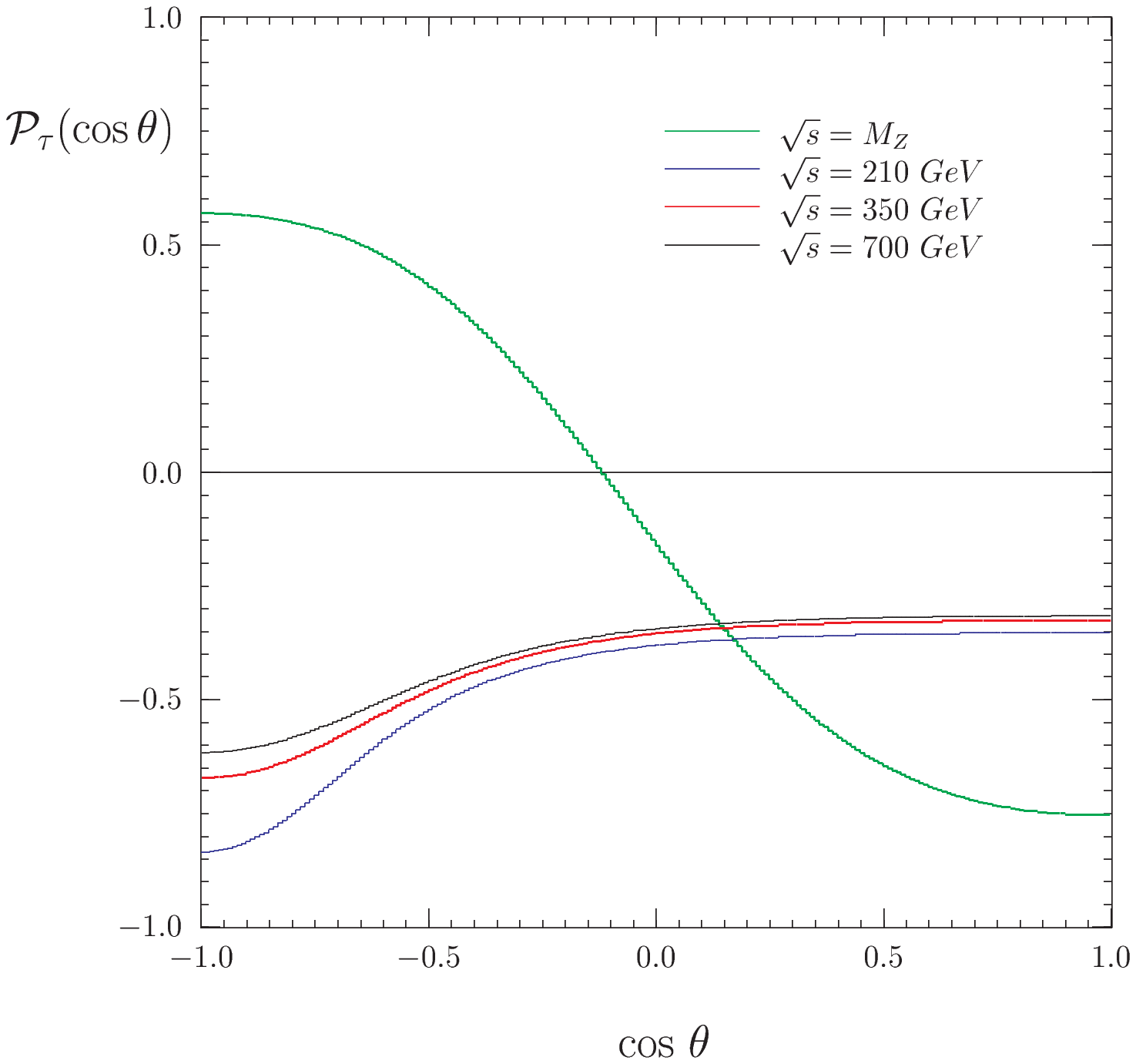,width=80mm,height=100mm}}}
\put(700, -220){\makebox(0,0)[lb]{\epsfig{file=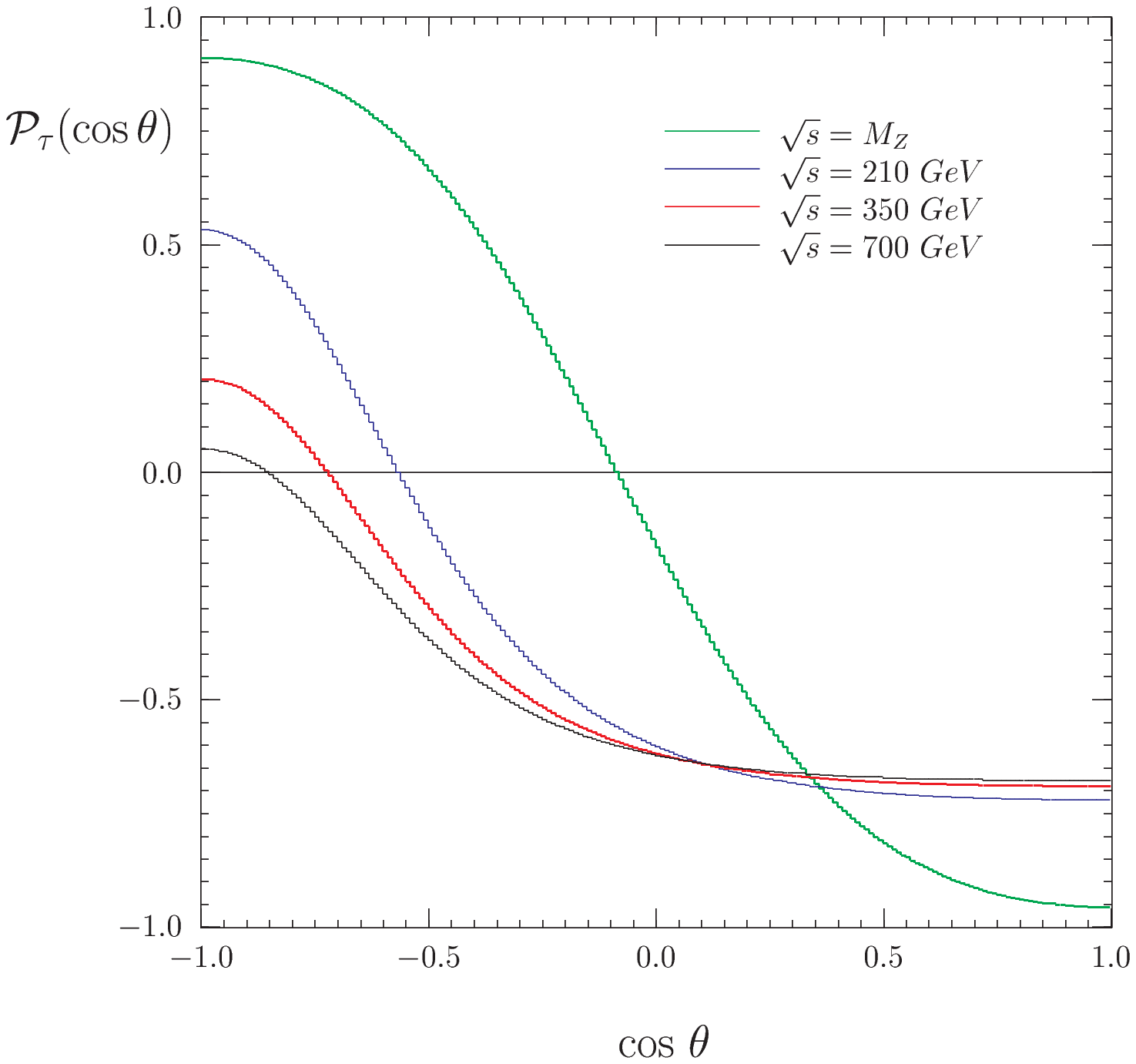,width=80mm,height=100mm}}}
\end{picture}
\caption
{\it Tests of the {\tt TAUOLA} universal interface. 
  The $\tau$ lepton  polarisation as a function }
\hspace{1.7cm}
{\it of $\cos\theta$. We have used 
$\sqrt{s}= M_Z$, 210, 350 and 700 GeV. }
\label{rozk2}
\end{figure}

In Fig.~\ref{rozk2}, the angular dependence of the $\tau$ polarisation
for the $e^{+}e^{-} \to \tau^+ \tau^-$ process
at the peak of $Z$ resonanse and for the three different values of the 
centre-of-mass (cms) energies above it (upper plot). 
The $\tau$'s are strongly polarised in the forward direction, 
where polarisation approaches the value approximately twice as big as 
the average one (the polarisation of Z due to $e-e-Z$ coupling sums with the
one due to $\tau-\tau-Z$ coupling).
The polarisation is very small in the backward regions.
For $\cos \theta=-1$ it is equal to zero, independently of the 
centre of mass energy. 
The reason is the universality of both the $Z$ and $\gamma$ couplings
to all leptons. 
As can be observed, the  $\tau$ polarisation changes significantly 
(especially in the forward region) with the
centre of mass energy. 
At cms  energies above the $Z$-peak,
$\tau$ polarisation is smaller, due to significant contribution from the  
$s$-channel $\gamma$-exchange -- production mechanism not contributing 
directly to the polarisation.

Let us now turn to the production from quarks.
In Fig.~\ref{rozk2} we show the $\tau$ polarisation for 
the $u\bar{u}\to \tau^+ \tau^-$ and  $d\bar{d}\to \tau^+ \tau^-$ elementary 
processes and the same cms energies as in the previous case. For the cms energy
close to the $Z$-peak,
the $\tau$'s produced in the forward as well as in the backward directions 
are strongly polarised. The polarisation seem to be almost zero for 
$\cos\theta=0$, but in reality is equal (nearly) to the same value as 
in previous case of production from electrons at $\cos\theta=0$.
Also, the average $\tau$ polarisation is close to the case of production from 
electrons. It is, as it should be, independent 
from the initial state flavour, the $Z$ was produced from. 
The initial state couplings of the $Z$,  affect  the angular dependence
of the $\tau$ polarisation only, and give sizable angular asymmetry 
in polarisation. Obviously, the larger the initial state vector
couplings to $Z$ the larger angular dependence of the polarisation.
The above arguments hold, because the contribution
from the $\gamma$ exchange is small. This is the case,  
in the region of the $Z$ peak only.

Quite different  polarisation pattern can be observed for quarks and 
cms energies far above the $Z$ peak (see Fig.~\ref{rozk2}).
Contribution from the $\gamma$ exchange
cannot be neglected. 
The $\gamma-Z$ interference  complicates the pattern even more.
Let us comment briefly on the numerical results.
In the case of $u$ and $d$  initial 
state, polarisation is negative and nearly constant over the
 forward hemisphere;
also the average polarisation is negative.
This is because of rather large and positive forward-backward asymmetry, 
and polarisation forward-backward asymmetry.
The average polarisation increases above  the Z-peak and 
approaches, respectively,  -35\% and -61\% 
in the case of $u \bar u$ and     
$d \bar d$ annihilation\footnote{This may open the way for measuring 
the flavour of the quarks leading to $\tau$ pair production.}. 
Let us point, that in case of  the $\tau$ production from the $d$ quarks
in the ultra high-energy limit, polarisation of the $\tau$ leptons in 
the backward directions approaches zero (independently of the numerical value of 
$\sin^2\theta_W$).

\section{ Towards spin sensitive observables}

Once we have understood the pattern of the $\tau$ polarisation, 
let us turn to the question of its measurement. 
We will follow the reasoning similar to the one of Ref.~\cite{cyt1}.
The $\tau$ polarisation ${\cal P}_{\tau}$ can be measured from the
energy distribution of its decay products. 
To simplify the discussion of spin effects,
the decays  $\tau \to \pi \nu $ were used only\footnote{
In case of the 3-body decays, such as $\tau \to e(\mu) \nu \bar \nu$ 
sensitivity
is smaller and depends on energy of leptons; in case of $\pi \to \rho \nu$ to
exploit sensitivity in full, the reconstruction of $\pi^0$ is necessary.}.
We will compare the cases of vector  $Z/\gamma^*$ and scalar neutral Higgs
 boson, produced in hadron collisions and decaying into pair of $\tau$ leptons.
The $\tau$ rest-frame can  not be accessed experimentally\footnote{ 
In some cases the invariant mass of  $Z/\gamma^*$ or Higgs,
can be reconstructed from total bilans
of the observed transverse energies for all tracks in the event.},
the spin effects can be seen, however, through 
the effects on $\pi^+$ $\pi^-$ energy distributions and correlations, 
observed/defined for the laboratory frame. 
\begin{figure}[!ht]
\setlength{\unitlength}{0.1mm}
\begin{picture}(1600,800)
\put( 375,750){\makebox(0,0)[b]{\large }}
\put(1225,750){\makebox(0,0)[b]{\large }}
\put(-20, -1){\makebox(0,0)[lb]{\epsfig{file=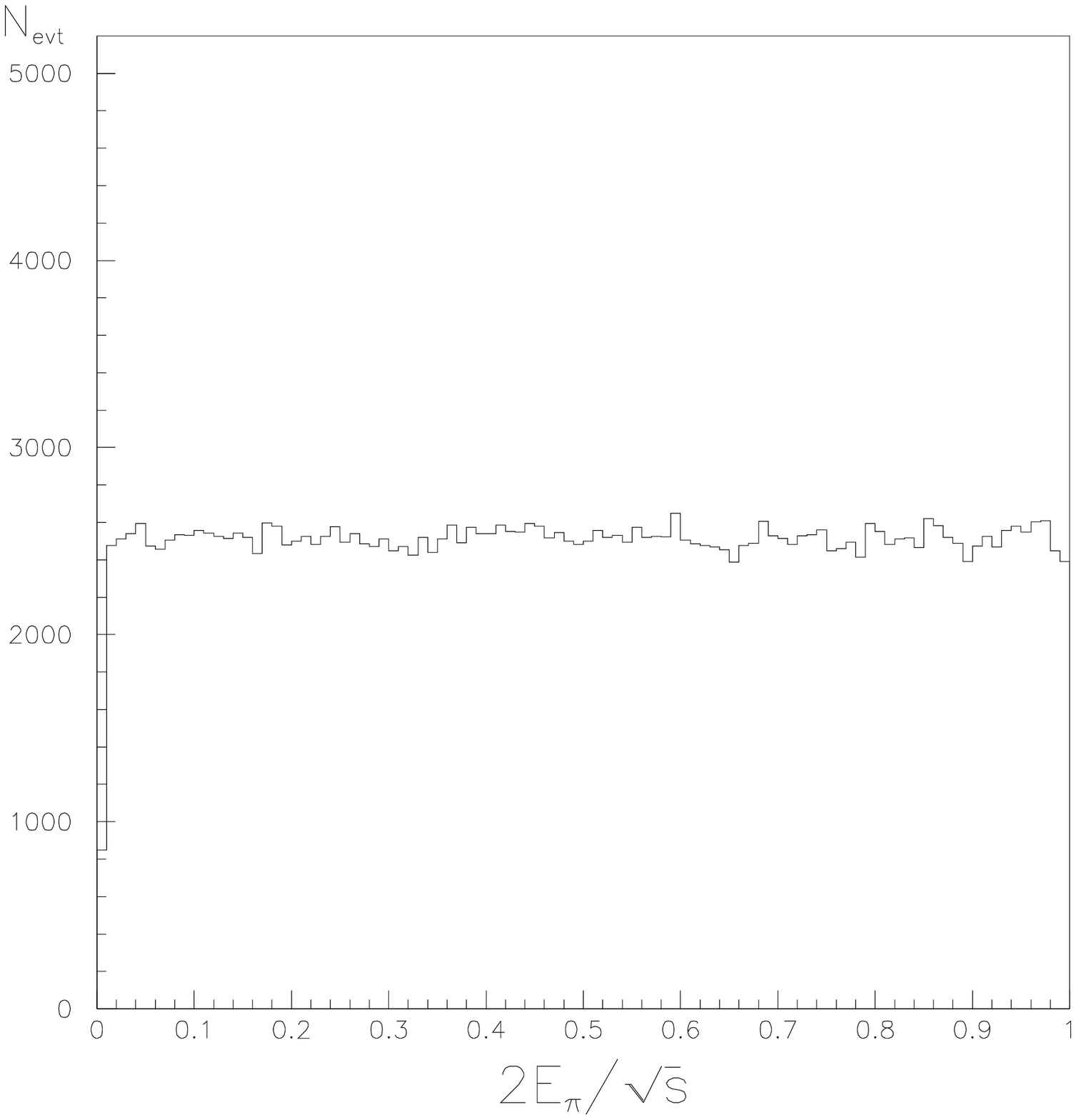,width=80mm,height=80mm}}}
\put(700, -1){\makebox(0,0)[lb]{\epsfig{file=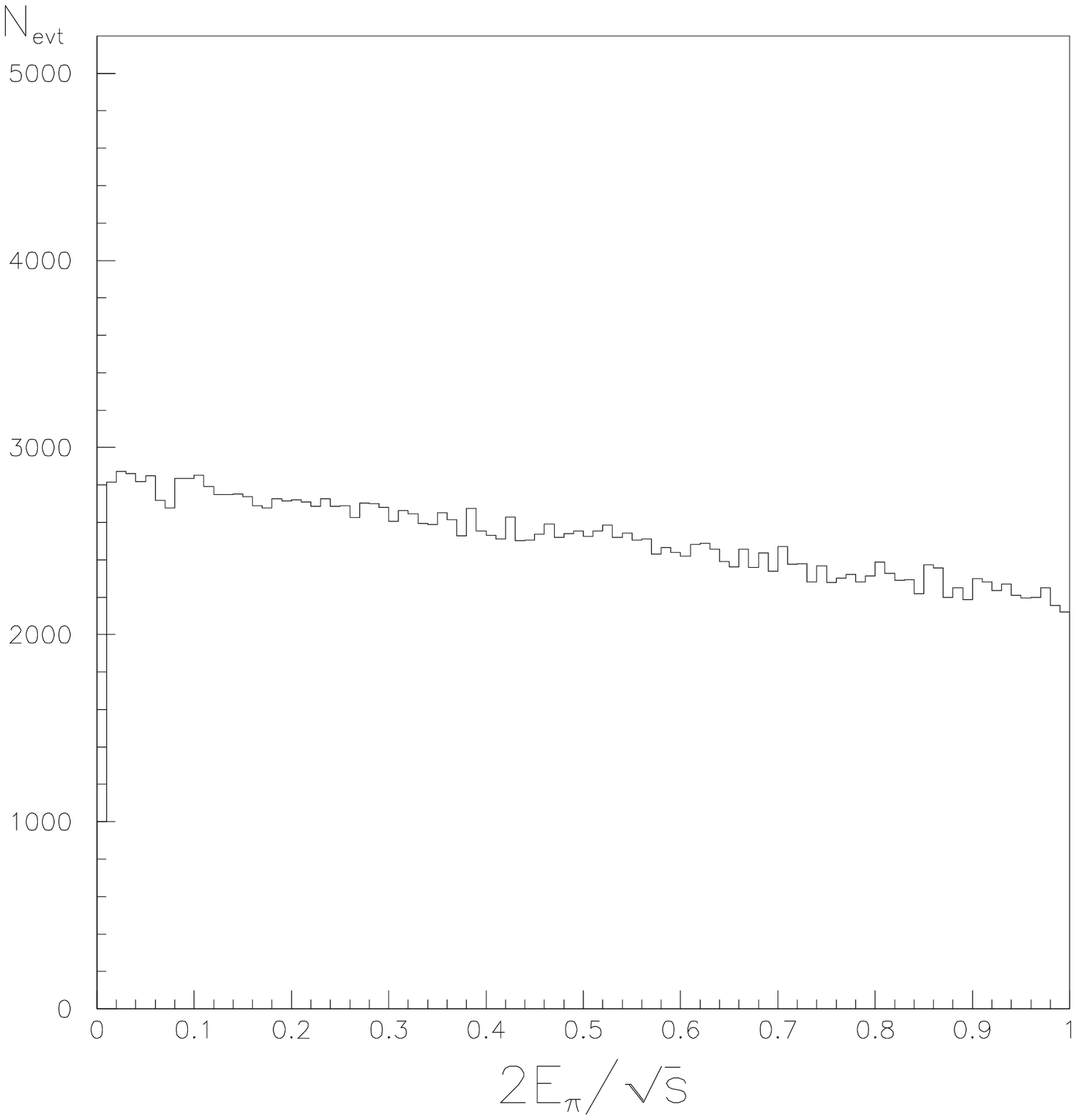,width=80mm,height=80mm}}}

\end{picture}
\caption
{\it Single $\pi$ energy spectrum in  the case of $\tau$ produced from $H$ 
(left-hand  side)  or $Z$ (right-hand  side).
$\sqrt{s} = m_{H}$ or $\sqrt{s} = m_{Z}$ respectively. Energies are calculated
in the rest frame of Higgs boson (or $Z$).}
\label{pion}
\end{figure}
  
Let us start  
with the following example, where for the production of the 
 $\tau$ lepton pairs
Monte Carlo program {\tt PYTHIA} was used, and for the decay Monte Carlo 
program {\tt TAUOLA}, and our interface.
It was assured, that the invariant mass of the pair of two incoming quarks
was $\sqrt{s}=m_{Z}= m_{H}$.  For the time being we discuss energies
defined in the $\tau^+\tau^-$ pair rest-frame.
With the help of variables $z_\pm= 2E_{\pi^\pm}/\sqrt{s}$, the spin 
effects are visualized.
In Fig. \ref{pion} we observe the  slope (as expected) of $\pi$ energy 
spectrum due to 
$\tau$ polarisation. The slope of the distribution is simply proportional 
to the polarisation (small dip at $z_\pm\simeq 0$ is due to kinematical effect
of the $\pi$ mass). 
\begin{equation}
{d\sigma \over d z_\pm} \sim 1 +{\cal P}_\tau \; 2\; (z_\pm-0.5)
\end{equation}
In the case of the plot on the right-hand side of Fig.~\ref{pion},
{\it i.e.} decay of the $Z$, polarisation was 
about -14.7\%. In the case of the plot on the left-hand side the 
spectrum is flat,
as would be in the case of scalar neutral Higgs boson (or pure $\gamma$) where
there is no polarisation.

\begin{figure}[!ht]
\setlength{\unitlength}{0.1mm}
\begin{picture}(1600,800)
\put( 375,750){\makebox(0,0)[b]{\large }}
\put(1225,750){\makebox(0,0)[b]{\large }}
\put(-20, -1){\makebox(0,0)[lb]{\epsfig{file=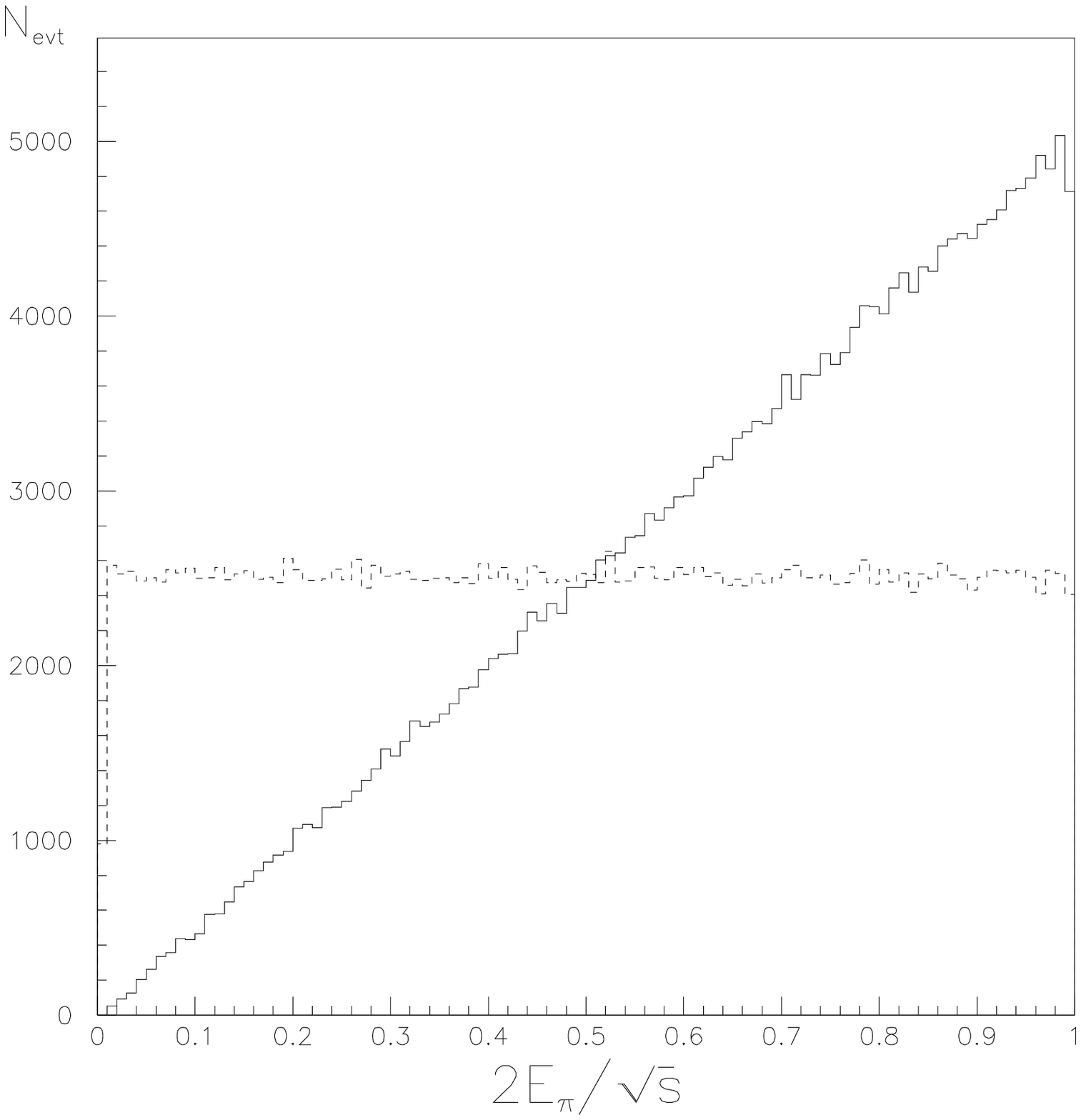,width=80mm,height=80mm}}}
\put(700, -1){\makebox(0,0)[lb]{\epsfig{file=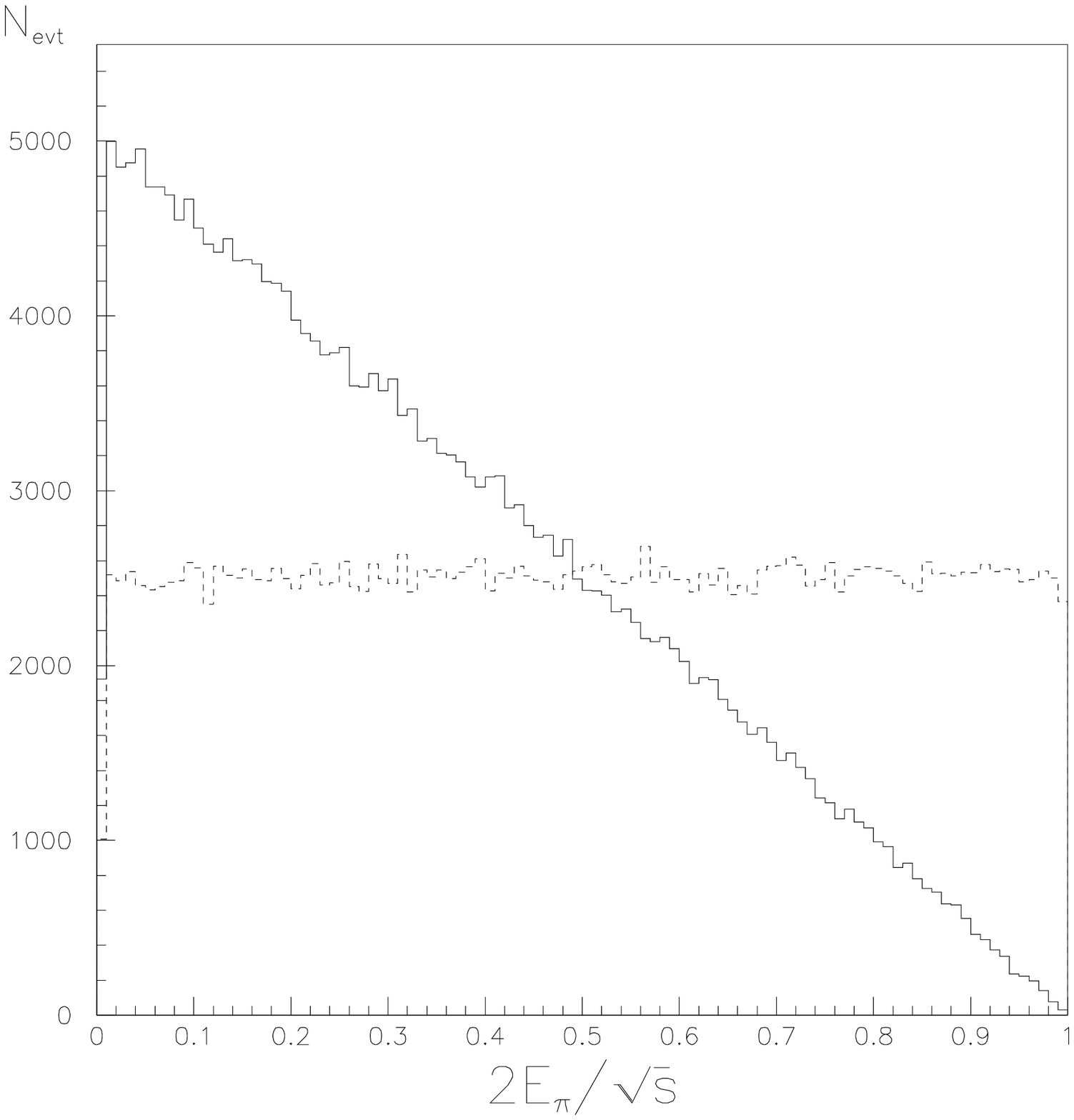,width=80mm,height=80mm}}}
\end{picture}
\caption
{\it Single $\pi$ energy spectrum in  the case of $\tau$ produced from $H^{\pm}$ 
(left-hand  side)  or $W^{\pm}$ (right-hand  side).
$\sqrt{s} = m_{H^{\pm}}$ or $\sqrt{s} = m_{W^{\pm}}$ respectively. 
Energies are calculated in the rest frame of $H^{\pm}$ (or $W^{\pm}$). 
Continuous line with spin effects included, dotted line with spin  effects 
switched off. }
\label{pion-charged}
\end{figure}

If polarisation was ${\cal P}_\tau$~=~$-$~100\%,
then the distribution slope would be maximally negative and touch  
zero at $z_\pm=1$. For ${\cal P}_\tau$~=~100\%, slope  would be reversed
and distribution would touch zero at $z_\pm=0$.
These are the cases of charged Higgs boson and charged W boson decays into 
 $\tau \nu$. Respective distributions are shown in Fig.~\ref{pion-charged}.

Let us now turn to the question of the spin correlations. As we can see
from Table~\ref{T:Probability}, the $\tau$ pairs are produced
with the well defined spin configurations ( ${+,+}$ or ${-,-}$
 for vector bosons;
 ${+,-}$ or  ${-,+}$ for neutral Higgs boson), thus, the spin effects are 
to become visible on 
the two-dimensional distribution build on $ z_-$ and $z_+$ variables.
Indeed, for the vector bosons we expect more events when both 
$\pi^+$ and $\pi^-$ are on the upper side (case I)
($z_+ > 0.5$, $z_- > 0.5$  ) or lower side
(case II) ($z_+ < 0.5$, $z_- < 0.5$  ), with respect to the mixed 
ones:
$z_+ > 0.5$, $z_- < 0.5$ (case III) and $z_+ < 0.5$, $z_- > 0.5$  (case IV).

The appropriate asymmetry is:
\begin{equation}
A_{FastSlow}= {\sigma_{I}+\sigma_{II}-\sigma_{III}-\sigma_{IV}
 \over \sigma_{I}+\sigma_{II}+\sigma_{III}+\sigma_{IV}}=0.24.
\end{equation}
This can be compared  to the Higgs boson case, where asymmetry 
$ A_{FastSlow}=-0.25$.
For the Higgs boson case the  $ A_{FastSlow}$ is insensitive on 
the choice  of cms energy, as there is no interference effects at all.
The sign difference  is due to the 
the vector boson nature of  $Z$, as opposed to scalar nature of Higgs boson.

\begin{figure}[!ht]
\setlength{\unitlength}{0.1mm}
\begin{picture}(1600,800)
\put( 375,750){\makebox(0,0)[b]{\large }}
\put(1225,750){\makebox(0,0)[b]{\large }}
\put(-20, -1){\makebox(0,0)[lb]{\epsfig{file=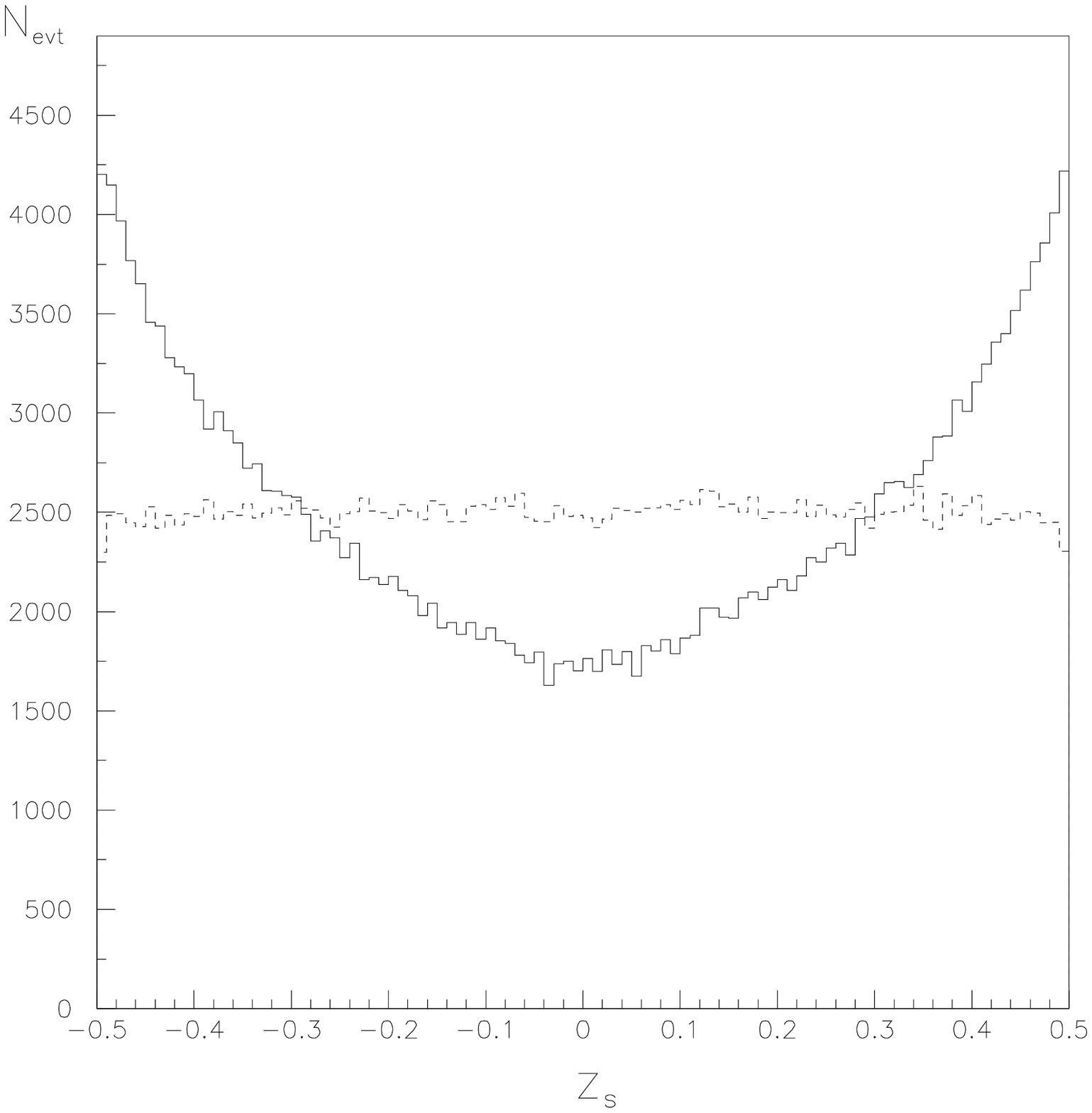,width=80mm,height=80mm}}}
\put(700, -1){\makebox(0,0)[lb]{\epsfig{file=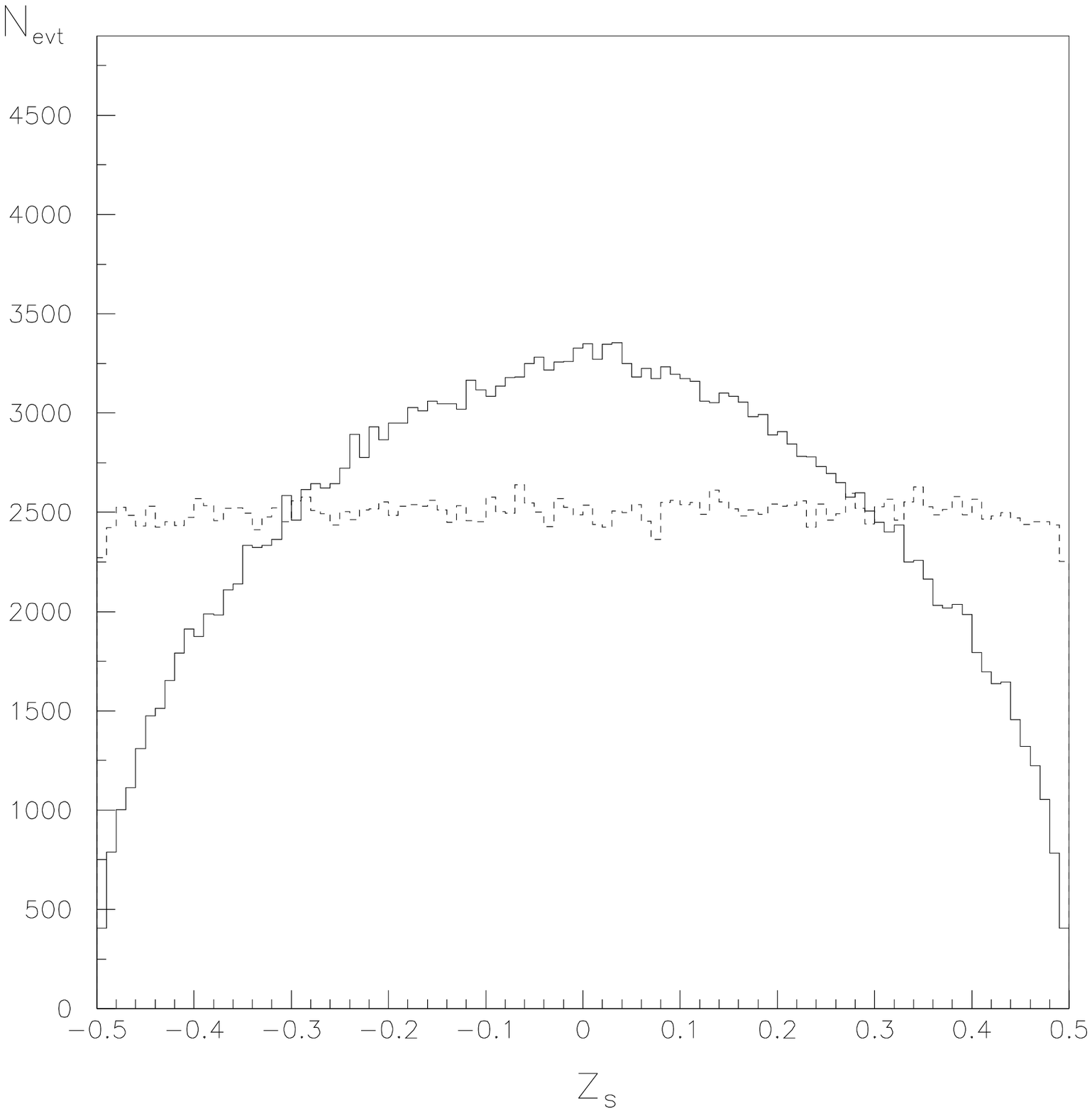,width=80mm,height=80mm}}}
\end{picture}
\caption
{\it Expected number of events as the function of $z_s$ 
variable. Left-hand  side plot for $H$, right-hand side for $Z$. 
Continuous line with spin effects included, dotted line with spin  effects 
switched off.}
\label{tomek}
\end{figure}

In order to  better visualize the spin correlation effect we have introduced 
variable $z_s$ defined as signed part of the following phase space part:
surface in $z_+$, $z_-$ variables between lines 
$z_+ =z_-$ and $z_+ =z_- +a$ 
(the sign of the  $a $ should be taken).
In  Fig.~\ref{tomek} respective plots are given for
the $H$ and $Z$ decays. 
The dashed lines (which in both $Z$ and $H$ cases are flat) correspond
to the case when spin correlations are switched off. 
As we expect
in the $Z/\gamma \rightarrow\tau^{-}\tau^{+}$ decays, due to spin effects a 
Fast (Slow) $\pi^{\pm}$ is most likely associated with a Fast (Slow)
$\pi^{\mp}$. The solid line has maximum at $z_s=0$
and approaches zero for $z_s=\pm 0.5$. Precisely the opposite is true in 
the case of  
$H\rightarrow\tau^{-}\tau^{+}$.  The maximum is now at $z_s=\pm 0.5$
and minimum for $z_s=0$.

\begin{figure}[!ht]
\setlength{\unitlength}{0.1mm}
\begin{picture}(1600,800)
\put( 375,750){\makebox(0,0)[b]{\large }}
\put(1225,750){\makebox(0,0)[b]{\large }}
\put(400, -1){\makebox(0,0)[lb]{\epsfig{file=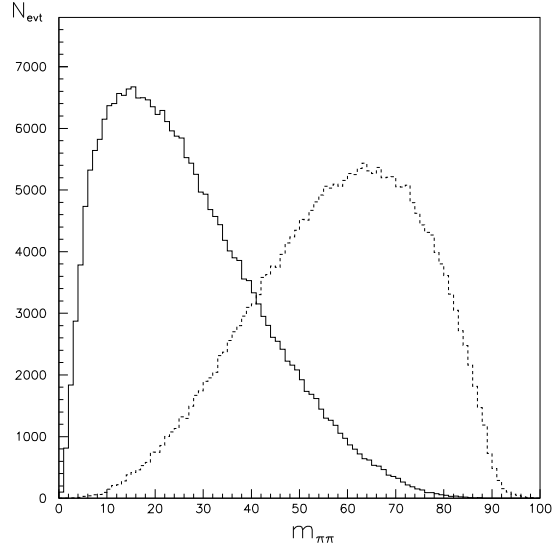,width=80mm,height=80mm}}}
\end{picture}
\caption
{\it  The $\pi^{+}\pi^{-}$ invariant
 mass distribution for $\sqrt{s}=m_Z$. Continuous line for pure $(-,-) $ 
 dotted line for pure $(+,+)$ spin configurations.}
\label{przod-tyl}
\end{figure}
\begin{figure}[!ht]
\setlength{\unitlength}{0.1mm}
\begin{picture}(1600,800)
\put( 375,750){\makebox(0,0)[b]{\large }}
\put(1225,750){\makebox(0,0)[b]{\large }}
\put(-20, -1){\makebox(0,0)[lb]{\epsfig{file=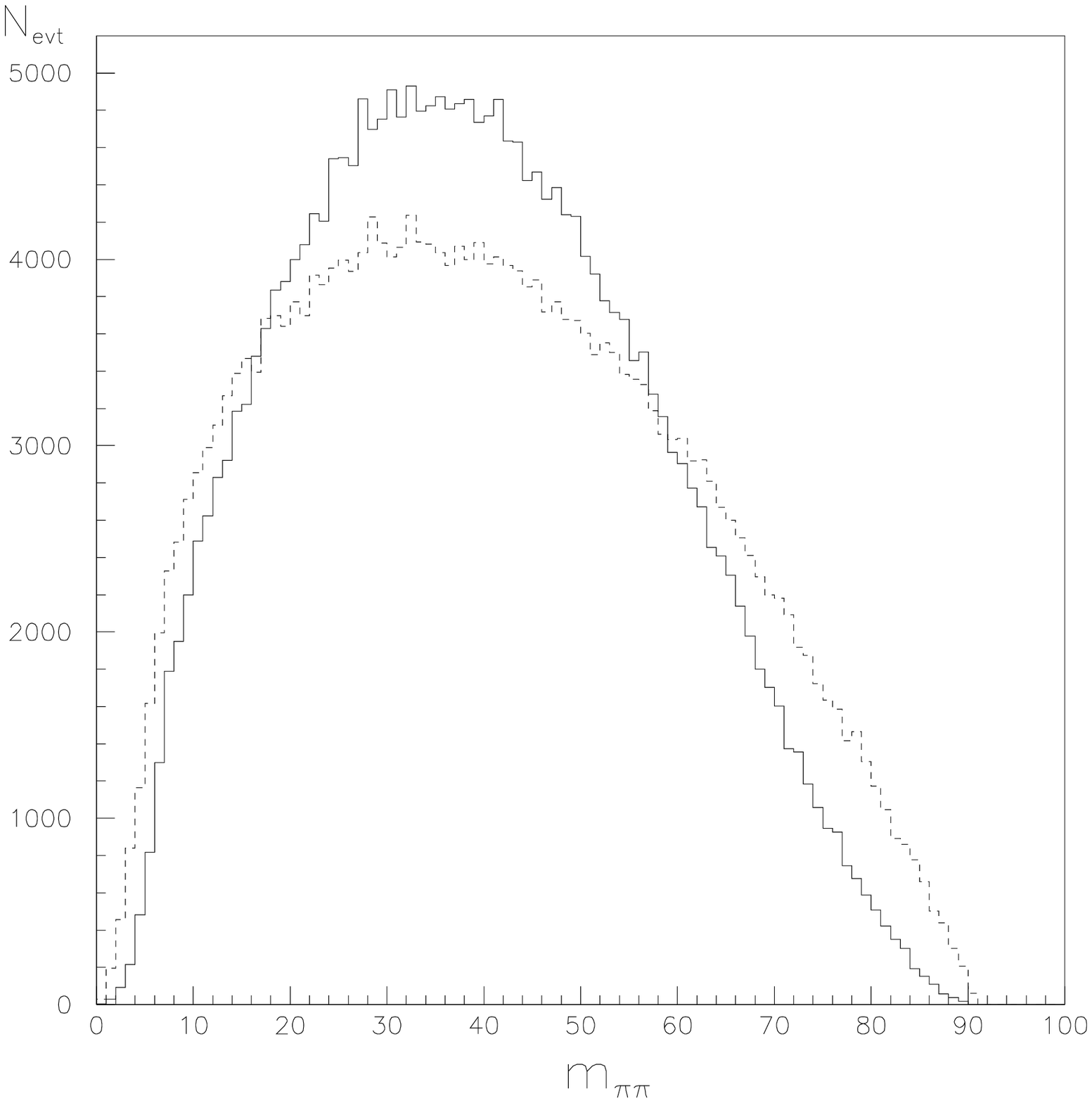,width=80mm,height=80mm}}}
\put(700, -1){\makebox(0,0)[lb]{\epsfig{file=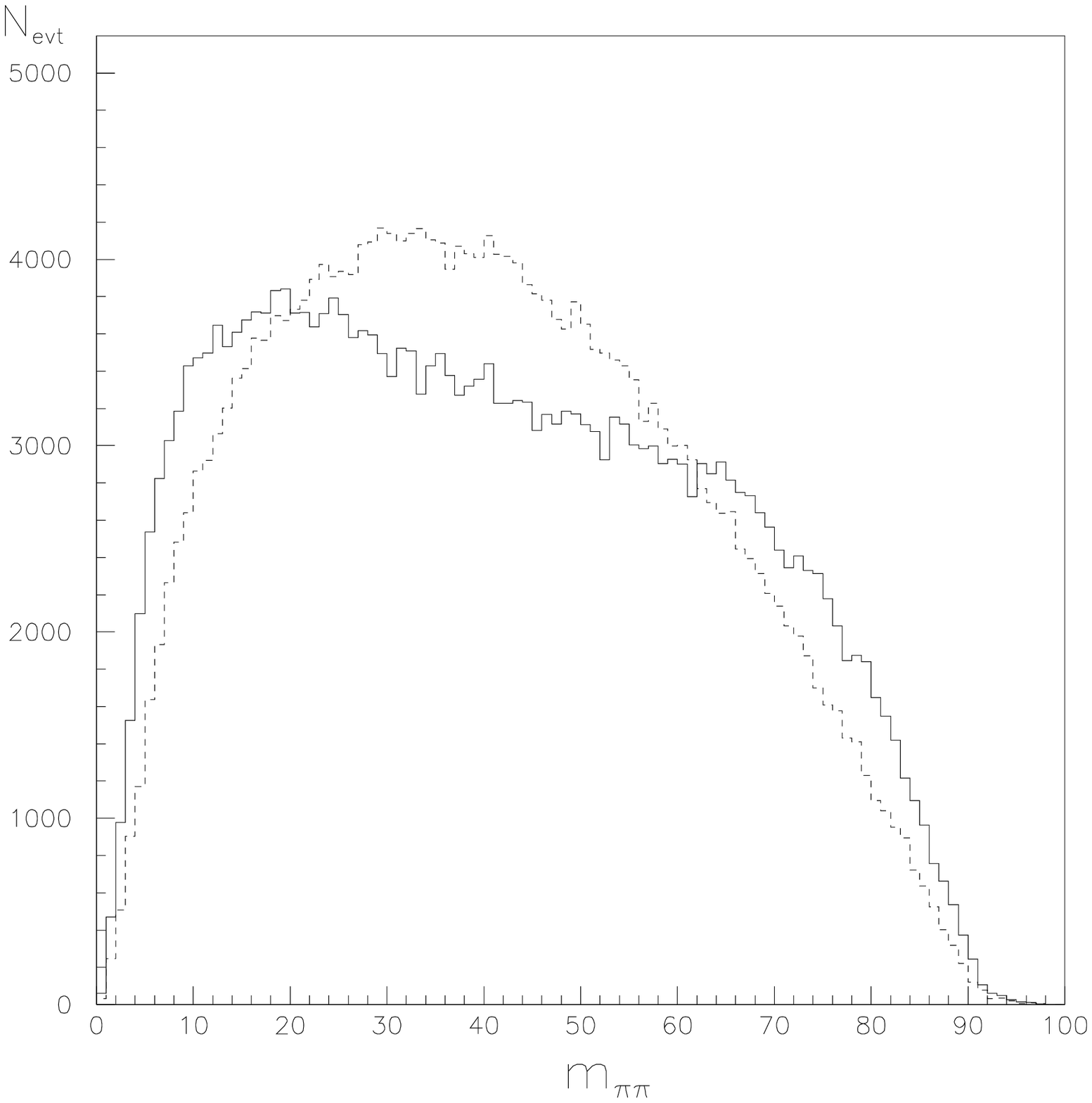,width=80mm,height=80mm}}}
\end{picture}
\caption 
{\it  The $\pi^{+}\pi^{-}$ invariant
 mass distribution. Left-hand side  plot for $H$; right-hand side for $Z/\gamma^*$.  
Continuous line with spin effects included, dotted line with spin 
effects switched off. In the two cases respectively $\sqrt{s}=m_H= m_Z$.   }
\label{invariantmass90}
\end{figure}

Let us now turn our attention 
to the quantities which (at least in principle) can be measured 
 experimentally.
Fig.~\ref{przod-tyl} shows the invariant mass 
distribution for the case of pure $(+,+)$ and $(-,-)$ configurations of the 
$\tau$ polarisation and the mass of the resonance equal to the $Z$ mass.
The spin correlations enhance fraction of the Fast-Fast and Slow-Slow configurations
which are localised mostly at the shoulders of the  $\pi^+\pi^-$
invariant mass distributions,
the  Fast-Slow configurations would be localised in the centre of the distributions.

In Fig.~\ref{invariantmass90} we show $\pi^+\pi^-$ invariant mass 
distribution for the  Higgs and $Z$ cases.
Continuous line with spin effects included, dotted line with spin 
effects switched off. Left-hand side  plot corresponds to the Higgs boson case,
right-hand side to the $Z$.
In the case of 
Higgs boson, the mass distribution is peaked centrally, whereas in the
 case of $Z/\gamma^*$
shoulders of the distributions are more profound, 
especially the one at lower invariant mass  additionally enhanced by the 
polarisation. 

For the vector case the spin correlations 
enhance fraction of the Fast-Fast and Slow-Slow configurations which are localised mostly 
at the shoulders of the  $\pi^+\pi^-$ invariant mass distributions.
The relative height of the shoulders is sensitive to the
average polarisation. For the scalar case, the enhancement in the fraction of the 
Slow-Fast configurations effects in the  enhancement of the events localised 
in the middle of the  $\pi^+\pi^-$ invariant mass distribution, leading to 
the slightly narrower shape.

If all polarisation effects are switched 
off (dashed lines) the distributions in the two cases are identical.
That observable,
well defined  distribution of  invariant mass built from  
the visible decay products of the $\tau$'s,
can be helpful in separating Higgs boson signal from $Z/\gamma^{*}$ background.

\begin{figure}[!ht]
\setlength{\unitlength}{0.1mm}
\begin{picture}(1600,800)
\put( 375,750){\makebox(0,0)[b]{\large }}
\put(1225,750){\makebox(0,0)[b]{\large }}
\put(-20, -1){\makebox(0,0)[lb]{\epsfig{file=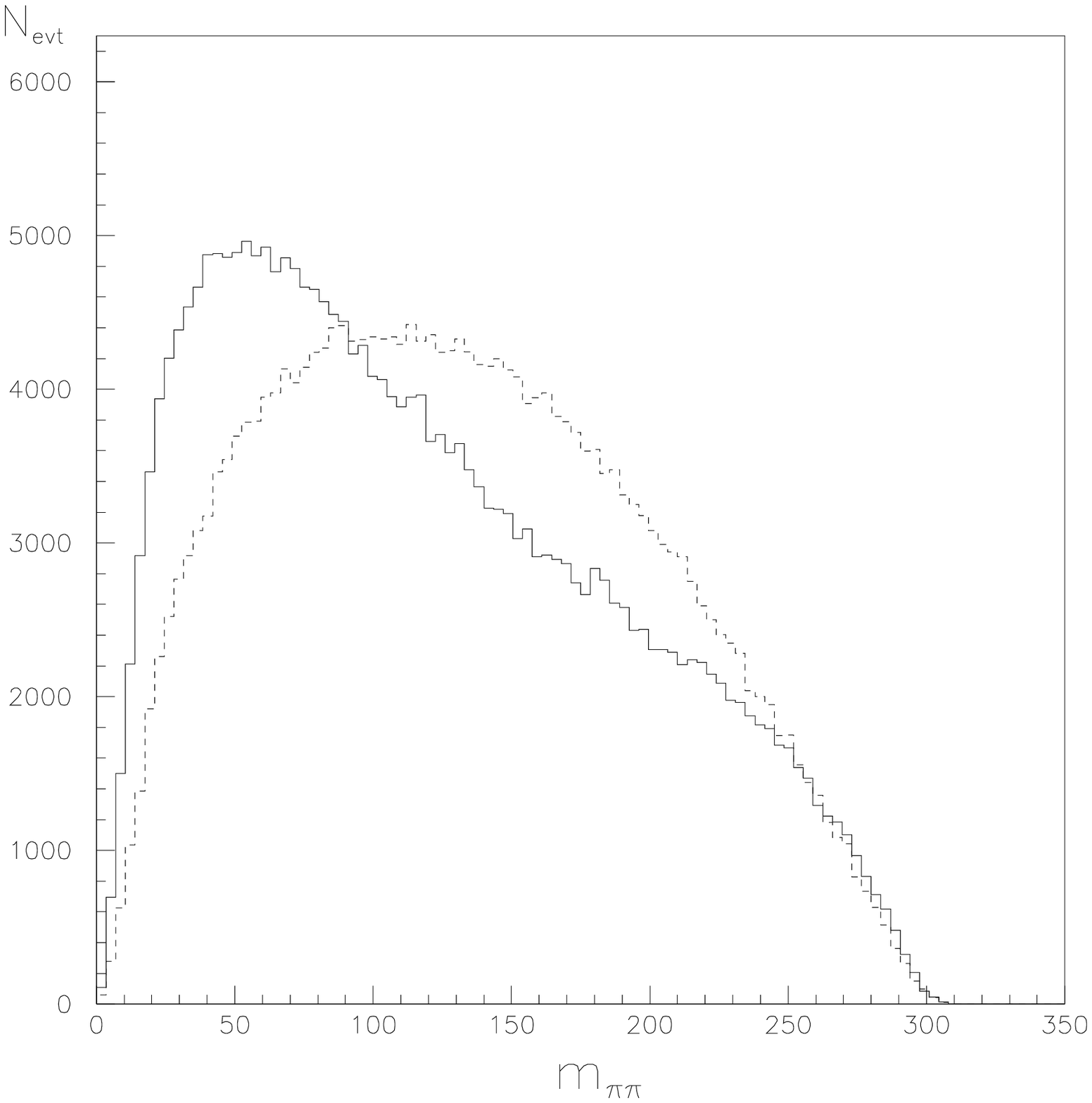,width=80mm,height=80mm}}}
\put(700, -1){\makebox(0,0)[lb]{\epsfig{file=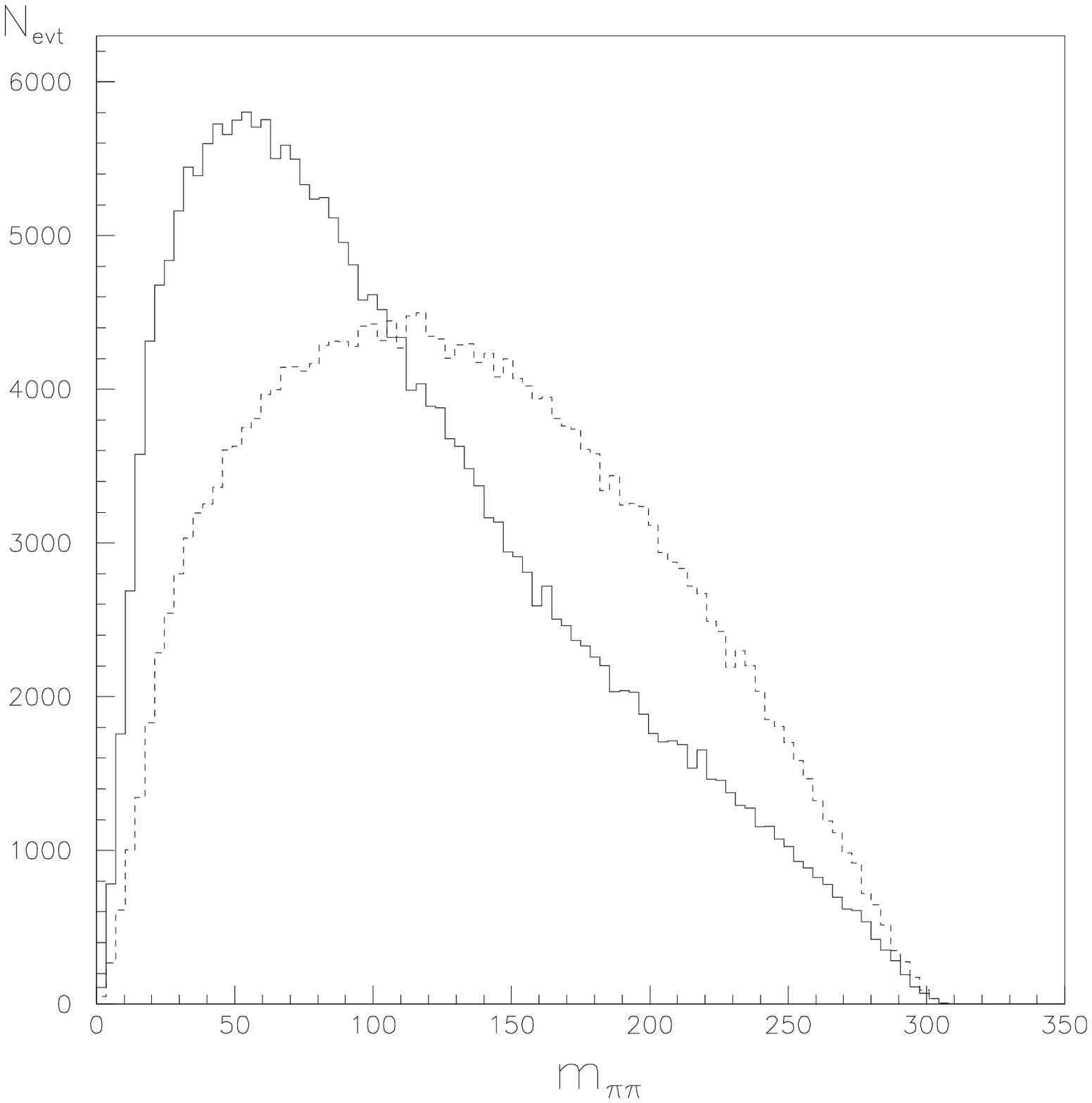,width=80mm,height=80mm}}}
\end{picture}
\caption 
{\it  The $\pi^{+}\pi^{-}$ invariant
 mass distribution for $u \bar u \to Z/\gamma^*$ (left-hand plot) and
$d \bar d \to Z/\gamma^*$ (right-hand plot) produced with cms energy of 300~GeV.  
Continuous line with spin effects included, dotted line with spin 
effects switched off.   }
\label{invariantmass300}
\end{figure}

The same distribution have been also studied for the off-peak production of 
$Z/\gamma^*$, {\it i.e.} for the larger cms energies. In these cases the 
average polarisation 
is large and negative, also distinct for the $u \bar u$ and $d \bar d$ annihilations.
As illustrated in Fig.~\ref{invariantmass300}, the effect on the 
$\pi^{+}\pi^{-}$ invariant mass distribution is noticeable.
The shape of the distribution might give the insight to the structure 
functions of the colliding protons, once the invariant mass of the $Z/\gamma^*$
state can be reconstructed.

\vskip 0.4 cm
\section{ Case of the Higgs signatures at LHC.}
\vskip 0.3 cm 
In the search for new phenomena in accelerator experiments,  important
parameter to estimate discovery chances is the signal sensitivity, equal to
the ratio of number of expected events from the new physics divided by the 
square root of the expected background events. 
Any possible increase of such a ratio can
improve chances of discovery (or improve limit of the exclusion). 
Use of more 
sophisticated cuts can be of a great help in a case when it can be combined 
with the physical properties of signal and/or background distributions.
Equally important is the possibility for the verification of the 
nature of observed new physics, {\it e.g.} the quantum numbers of 
the new resonances. 

The $\tau$ leptons are considered as a very promising signature for the searches
of the Higgs bosons in the Minimal Supersymmetric Standard Model (MSSM) 
at LHC collider \cite{ATLAS-TDR, CMS-TP}. 
Below, we will briefly discuss possible applications of the discussed spin correlations 
in the cases of the neutral Higgs bosons $H$ and $A$ decay 
into  $\tau^+ \tau^- $ pair and the charged Higgs boson $H^{\pm}$ decays 
into $\tau \nu $ pair.

The neutral Higgs bosons H and A decays into  $\tau^+ \tau^- $ pair, 
are enhanced
for the large values of $\tan \beta$
($\tan \beta$ denotes the ratio of the vacuum expectation values of the
 Higgs doublets in the MSSM model),
with the branching ratio of about 10\% for most of the range of the interesting 
Higgs boson mass values ( 150-1000 GeV ). Accessibility of the hadronic decay 
mode of the $\tau^+ \tau^- $ pair has been studied recently by the CMS Collaboration
\cite{CMS-1999-037}. The triggering on such events, signal extraction from the 
irreducible background  $Z/\gamma^* \to \tau \tau$ and the reducible 
backgrounds QCD jets, $t \bar t$ and $W+jets$, and reconstruction of the 
resonance peak in the  $\tau^+ \tau^- $ mass distributions seems feasible 
for the Higgs boson masses roughly above 300~GeV.
In that study the $\tau$ identification is based on the presence of a single hard 
isolated charged hadron in the jet using tracker information.
The two hard tracks from $\tau^{-}$
and $\tau^{+}$ in the signal events have an opposite sign while no strong charged 
correlation is expected for the QCD jets or  $W+jets$ events.
The sensitivity of 5$\sigma$ can be reached 
for the large fraction of the MSSM parameter space after three years of data 
collecting at low luminosity. The expected signal-to-background ratio is very large,
being of the order of one for 5$\sigma$ sensitivity, with the background dominated
by the continuum $Z/\gamma^* \to \tau \tau$ production. 
The resolution for the reconstruction of the gaussian $m_{\tau\tau}$
 peak is $\sim$ 
10\% of the mass of the Higgs boson. The above performance seems  
very promising but hopefully can be still improved by exploring the spin
correlations and polarisations effects. The possible improvements may come from the
additional suppression of the background.

 But it is also important that exploring
effect  of the spin correlation on the invariant mass of the hadronic decay products
might allow to verify hypothesis of the scalar {\it versus} vector nature  
of the observed 
resonance peak in the reconstructed invariant mass of the $\tau \tau$ pair.
As we have seen in the previous sections,
the main difference between production mechanisms due to
Higgs boson {\it versus} $Z/\gamma^{*}$ consists of the correlation in energies of the $\tau$ 
hadronic decay products.

\begin{figure}[!ht]
\centering
\setlength{\unitlength}{0.1mm}
\begin{picture}(1600,800)
\put( 375,750){\makebox(0,0)[b]{\large }}
\put(1225,750){\makebox(0,0)[b]{\large }}
\put(-20, -1){\makebox(0,0)[lb]{\epsfig{file=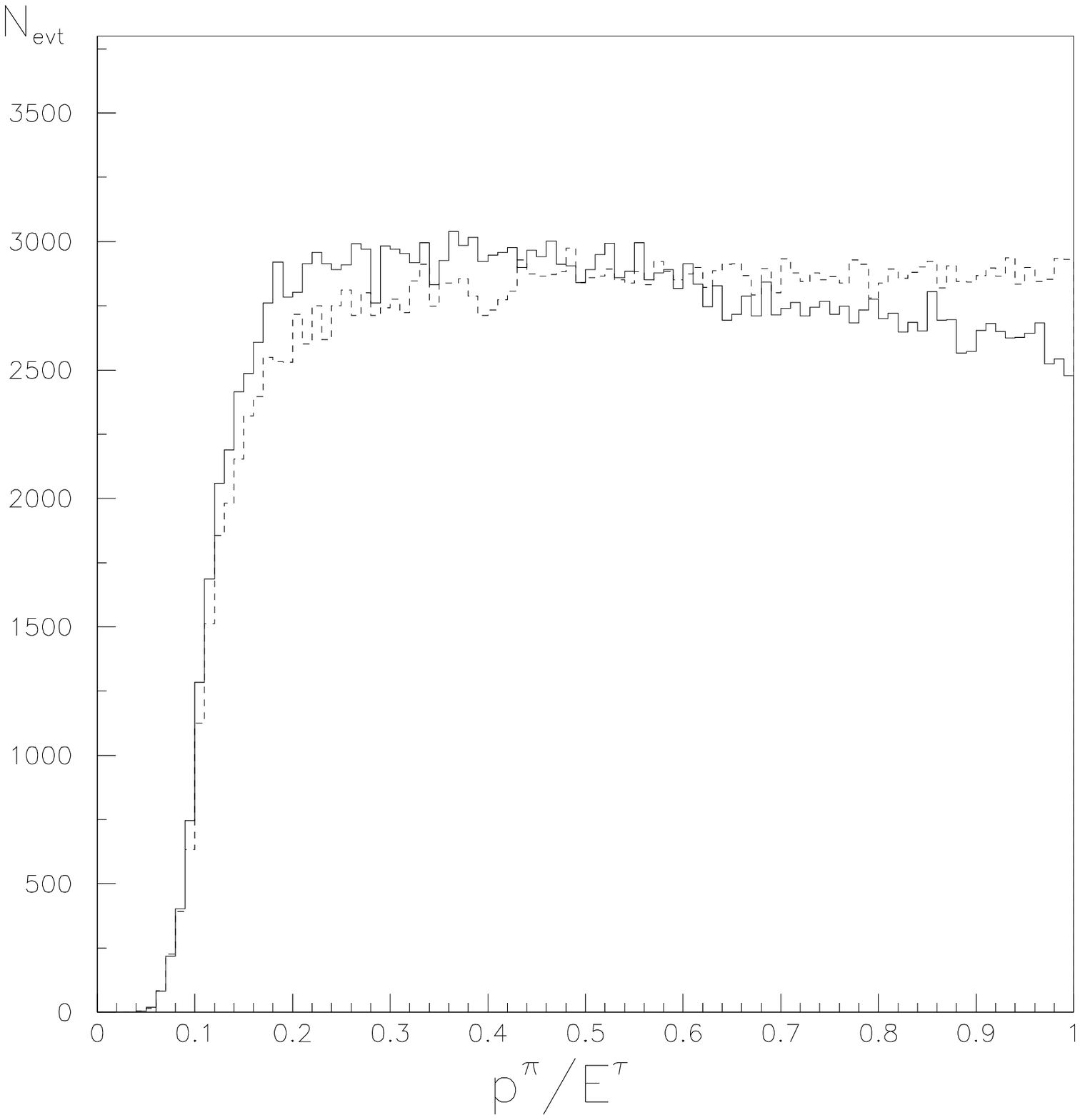,width=80mm,height=80mm}}}
\put(700, -1){\makebox(0,0)[lb]  {\epsfig{file=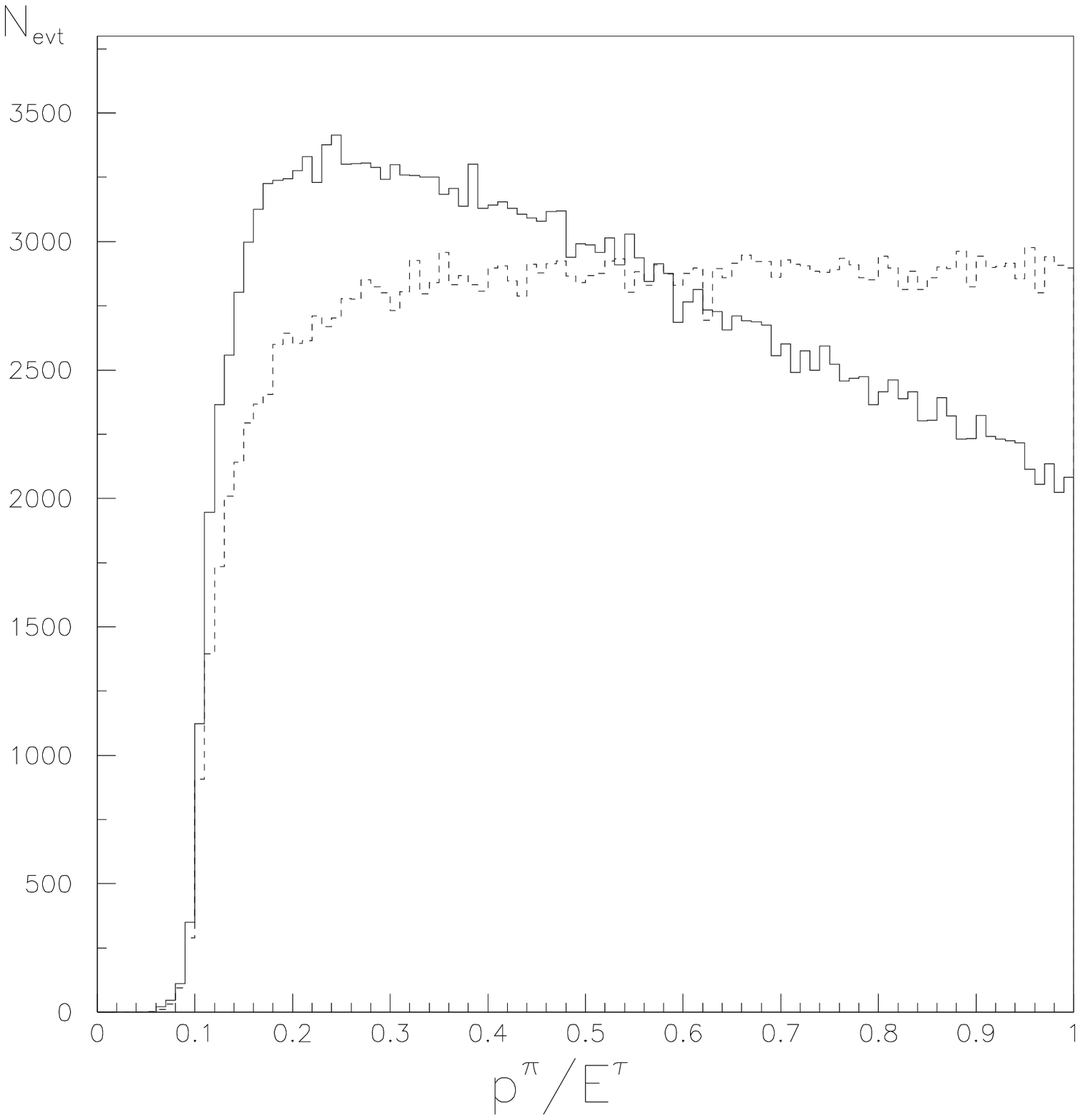,width=80mm,height=80mm}}}
\end{picture}
\caption
{\it 
The $\pi$ energy spectrum in the laboratory frame, 
after basic selection  as specified in text,
 for Higgs (left-hand side) and  $Z/\gamma^{*}$ (right-hand side).
Continuous line with spin effects included, dotted line with spin 
effects switched off.}
\label{cut1}
\end{figure}
\begin{figure}[!ht]
\centering
\setlength{\unitlength}{0.1mm}
\begin{picture}(1600,800)
\put( 375,750){\makebox(0,0)[b]{\large }}
\put(1225,750){\makebox(0,0)[b]{\large }}
\put(-20, -1){\makebox(0,0)[lb]{\epsfig{file=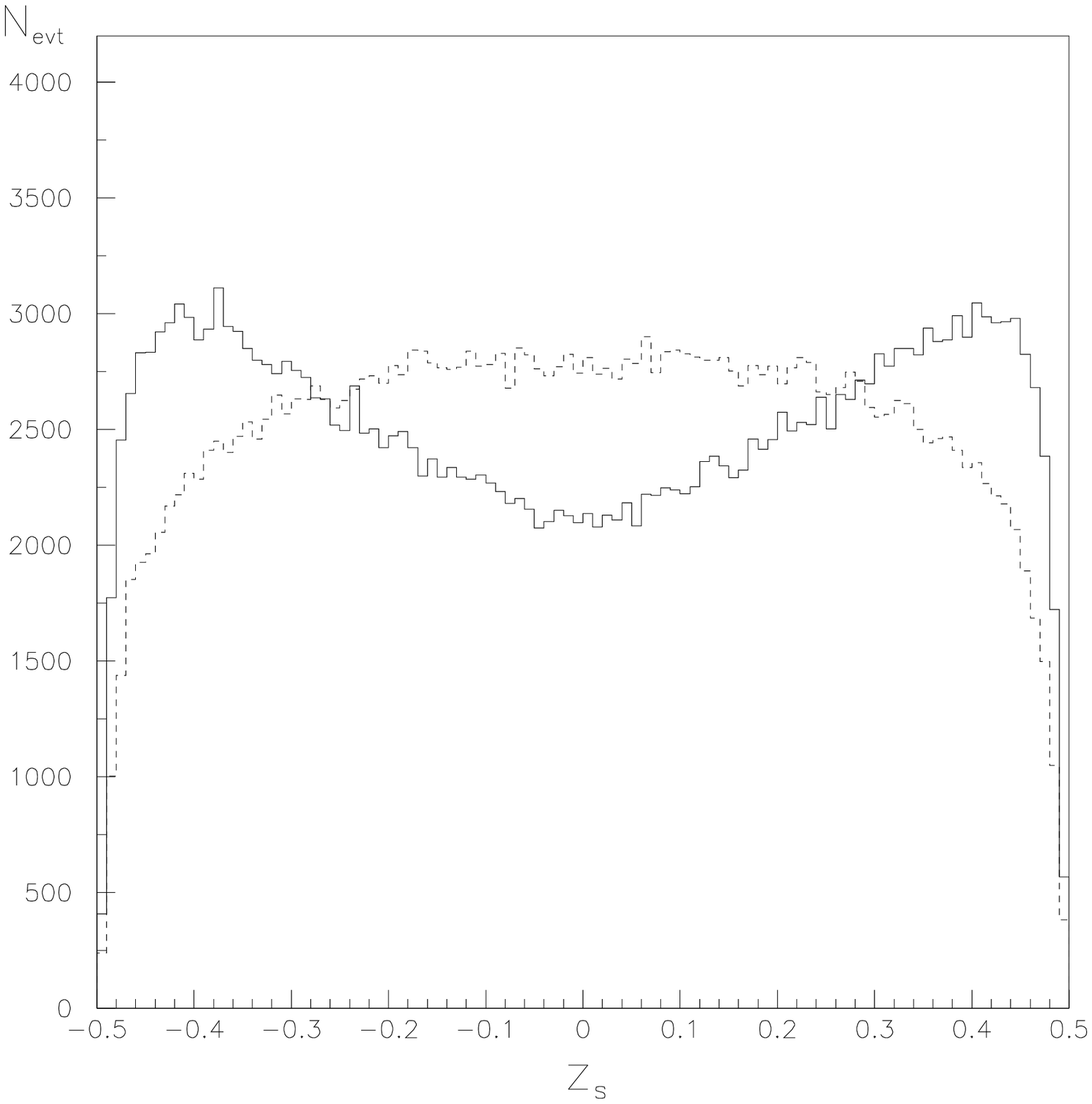,width=80mm,height=80mm}}}
\put(700, -1){\makebox(0,0)[lb]{\epsfig{file=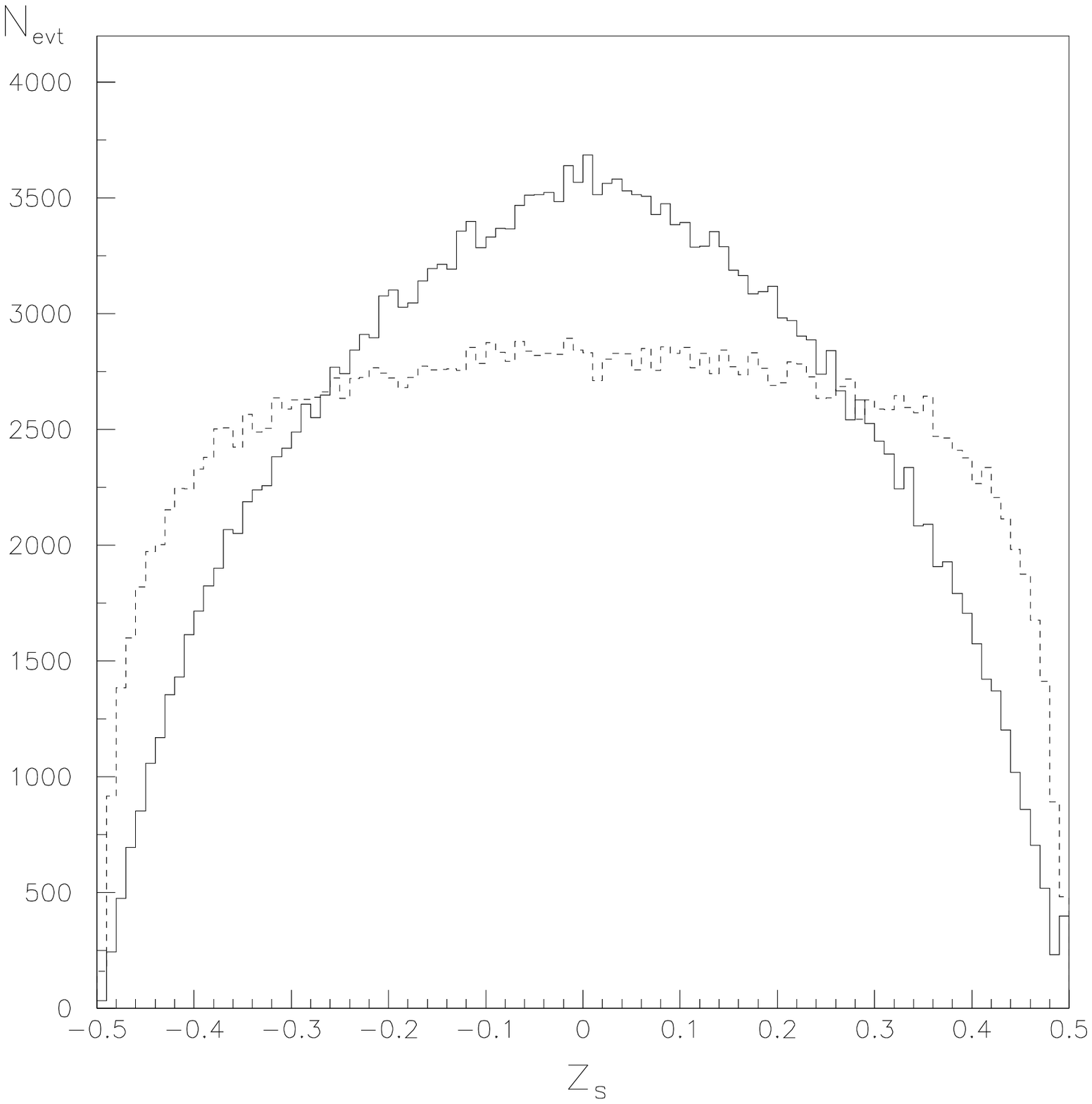,width=80mm,height=80mm}}}
\end{picture}
\caption
{\it The expected number of events as a function of the $\pi^{+}\pi^{-}$ 
Energy-Energy correlation variable $z_s$,  
basic selection as specified in text included. Distributions are shown for
Higgs  (left-hand side) and  $Z/\gamma^{*}$ (right-hand side).
Continuous line with spin effects included, dotted line with
 spin effects switched off.}
\label{cut4}
\end{figure}
Let us concentrate on the 
case of $\tau$ decays to $\pi\nu$ --most sensitive to the spin correlations.
In Fig.~\ref{cut1} the $\pi$ energy spectrum is shown in the laboratory
frame for the $H$ (left-hand side) and $Z/\gamma^*$ (right-hand side) decays. 
The selection, roughly consistent
with what is foreseen in the experimental analysis \cite{ATLAS-TDR}, was applied. 
The minimal transverse momenta of the  $\pi's$ were required to be above 15~GeV and 
the pseudorapidity $|\eta|~<~2.5$. 
Solid line shows results with included spin effects, dashed line with the
effects switched off. Spin correlations lead to
the softer spectrum of $\pi$ in $Z/\gamma^*$ decays, which may slightly 
suppress the identification efficiency of the  $\tau$ with respect to those produced from the Higgs boson decay.
The slope in the distribution for
the $Z/\gamma^*$ (sensitive to the large negative average polarisation),
depends also on the relative fraction of the $u \bar u$ and $d \bar d$ production
processes simulated in the proton-proton collision, 
hence the parametrisation of the structure functions.
Events were generated with the Monte Carlo {\tt PYTHIA 5.7}
 \cite{Sjostrand:2000wi}
and {\tt CTEQ2L} structure functions; the Higgs boson mass of the 300 GeV and the width below 1 GeV
(as for $\tan \beta~\sim~10$) was assumed. 
For the continuum $Z/\gamma^*$,
the cms energy of the produced $\tau \tau$ pair was taken in the 
range 300~$\pm$~10~GeV. 

Fig.~\ref{cut4} shows the effect of the $\pi \pi$ energy-energy correlations 
in the cms frame as discussed in the previous section.
Although the effect seems strong enough to discriminate between
scalar and vector cases, exploring this effect would require good experimental
reconstruction of the effective cms frame which might be very difficult.

\begin{figure}[!ht]
\centering
\setlength{\unitlength}{0.1mm}
\begin{picture}(1600,800)
\put( 375,750){\makebox(0,0)[b]{\large }}
\put(1225,750){\makebox(0,0)[b]{\large }}
\put(-20, -1){\makebox(0,0)[lb]{\epsfig{file=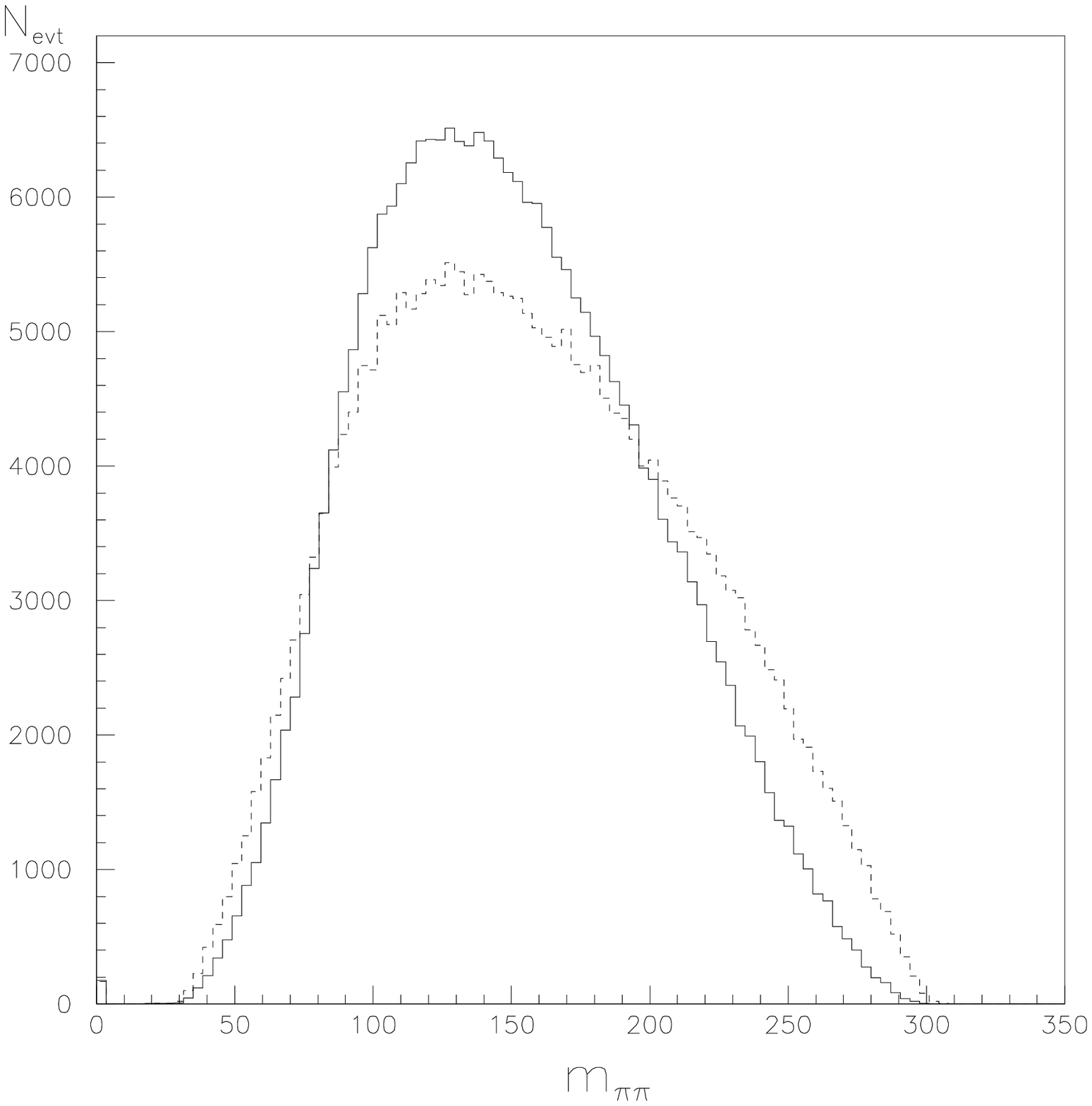,width=80mm,height=80mm}}}
\put(700, -1){\makebox(0,0)[lb]{\epsfig{file=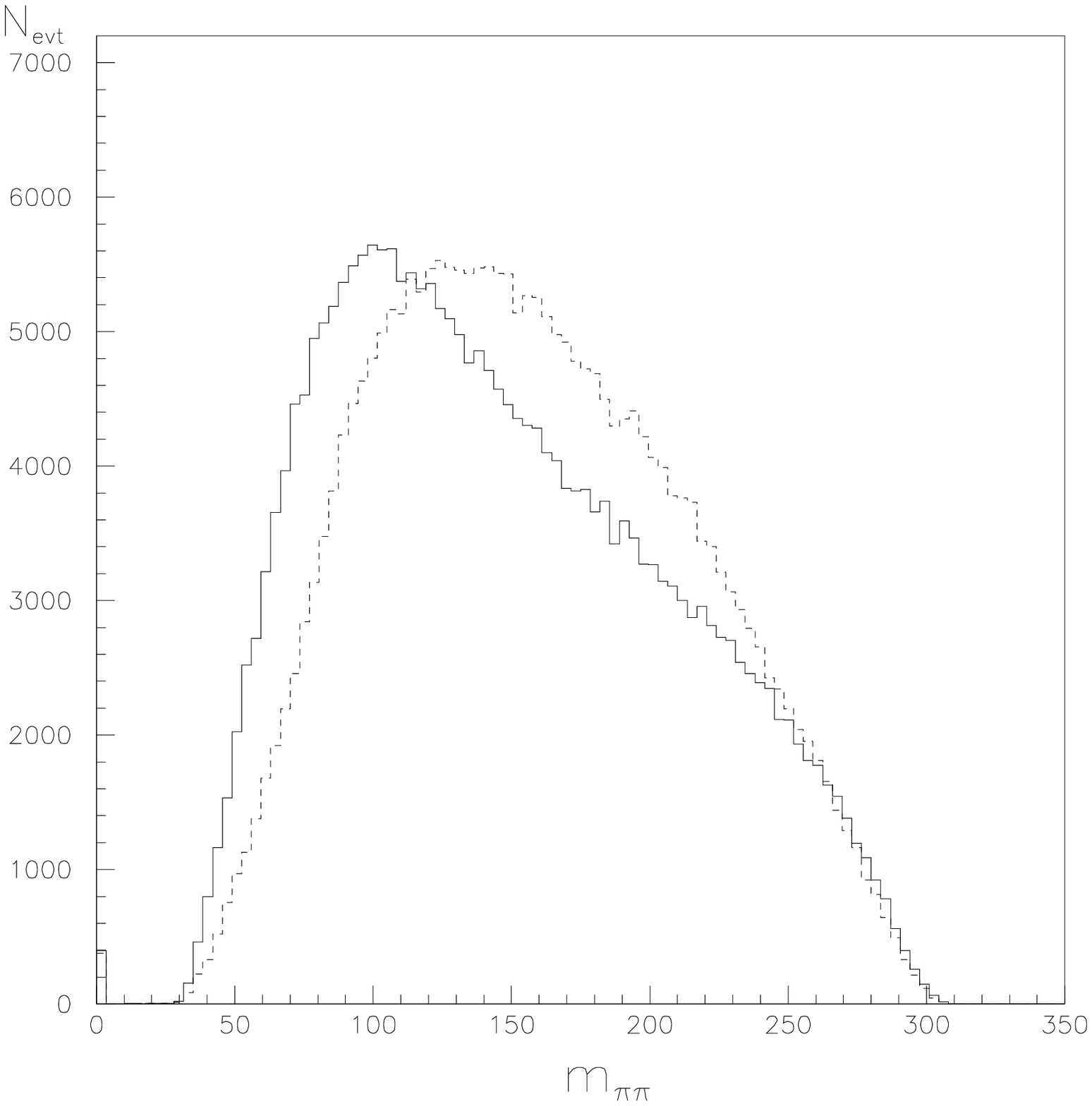,width=80mm,height=80mm}}}
\end{picture}
\caption
{\it The $\pi^{+}\pi^{-}$ invariant
 mass distribution after basic selection. 
On the left  for Higgs, on the right for $Z/\gamma^{*}$.
Continuous line with spin effects included, dotted line with spin 
effects switched off. The Higgs mass was assumed to be 300~GeV (see text).}
\label{cut5}
\end{figure}

The invariant mass distribution of the  $\pi \pi$ system, see Fig.~\ref{cut5},
shows the visible effect of including spin-correlations. 
As discussed in the previous
section, in the vector case these correlations lead to the more profound shoulders
in the distribution, and the relative height of them is sensitive
to the average polarisation of the produced resonance. In the case of the scalar Higgs boson
the correlations lead to the narrower 
distribution, enhancing fraction of events localised in its central part.
With the expected signal-to-background ratio being of one or higher, 
this effect seems promising for determining spin property 
of the studied resonance. 

For completeness let us now turn to the case of the charged Higgs boson. 
The decay into  $\tau \nu $ pair is a dominant mode
below the kinematical threshold of the $tb$ channel, 
and accounts for nearly 100\% of cases, almost independently of $\tan \beta$.
This branching ratio is decreasing
rather rapidly above $tb$ thereshold, but is still of the order of 10\% for the Higgs boson mass of 
500 GeV and large  $\tan \beta$. The LHC detectors will be thus sensitive to signal 
\cite{ATLAS-TDR, CMS-1999-037}, with at least $5\sigma$ significance, for  Higgs boson mass up to 400-500 GeV.
This sensitivity is almost independent on  $\tan \beta$ below  $tb$ threshold.
For the  Higgs boson mass above top-quark mass this sensitivity is expected only for large  $\tan \beta$.
In both cases of Higgs boson masses below and above top-quark mass, 
the main background comes from
  $W^{\pm}\rightarrow\tau^{\pm}\nu$ decay.
Harder pions are expected from the $H^{\pm}$ decay than from the $W^{\pm}$
decays.

\begin{figure}[!ht]
\centering
\setlength{\unitlength}{0.1mm}
\begin{picture}(1600,800)
\put( 375,750){\makebox(0,0)[b]{\large }}
\put(1225,750){\makebox(0,0)[b]{\large }}
\put(-20, -1){\makebox(0,0)[lb]{\epsfig{file=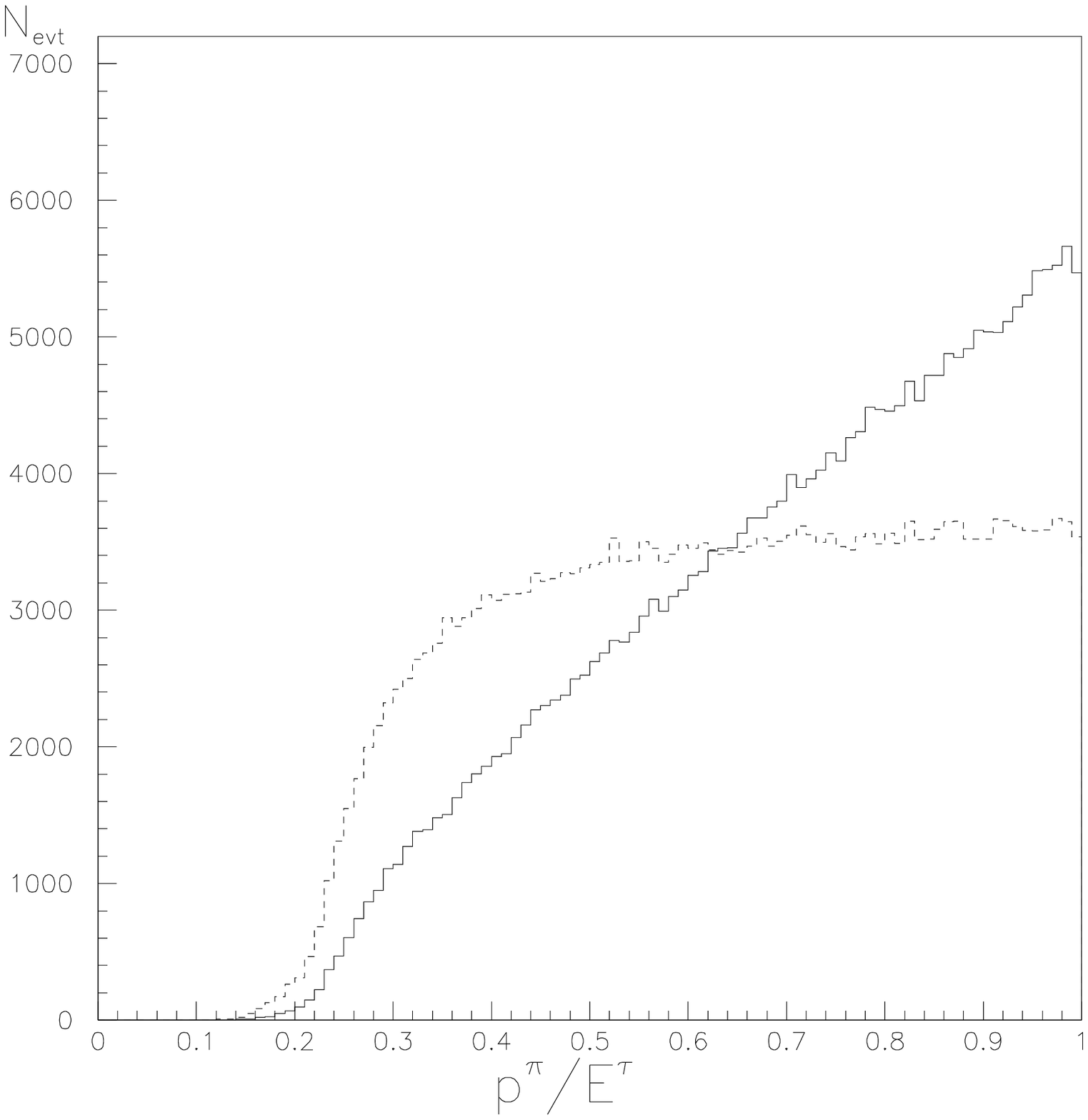,
width=80mm,height=80mm}}}
\put(700, -1){\makebox(0,0)[lb]{\epsfig{file=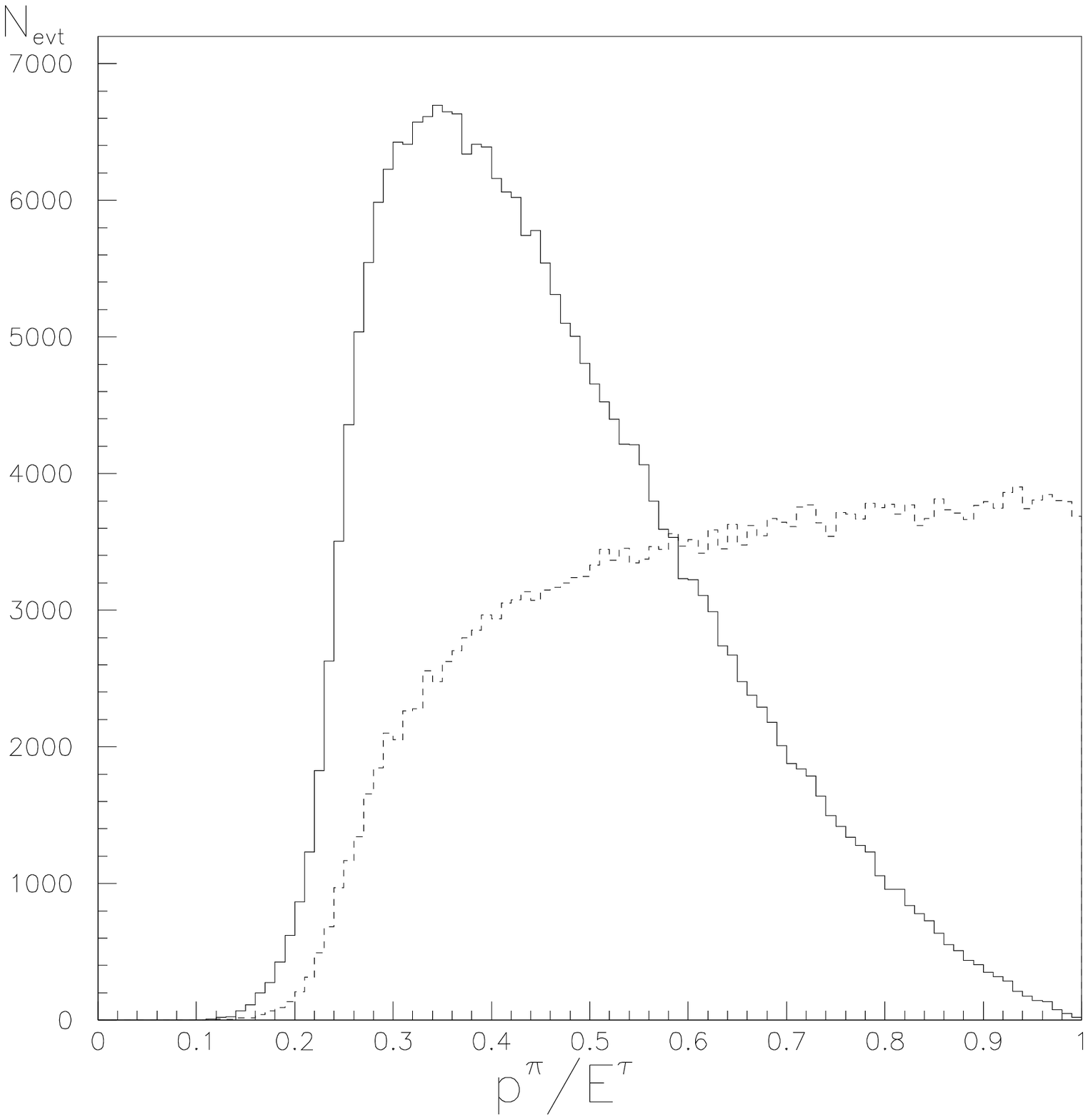,width=80mm,height=80mm}}}
\end{picture}
\caption
{\it 
Distribution of $p^{\pi}/E^{\tau}$ after basic selection and in the laboratory frame. On the left plot for $H^{\pm}$ ( $m_{H}=130$ $GeV$), on the right for $W^{\pm}$. 
Continuous line with spin effects included, dotted line with
 spin effects switched off. }
\label{cut6}
\end{figure}

The effect of the  spin correlations has been already studied, theoretically 
in Ref.~\cite{DPRoy} and experimentally in Ref.~\cite{CMS-1999-037}. 
For completeness we show in Fig.~\ref{cut6} the $\pi$ energy distribution
in the laboratory frame for the decays of $H^{\pm}$ and $W$ -- spin effects 
switched on and off. The Higgs boson mass of 130~GeV was taken. 

\newpage
\vskip 0.4 cm
\section{ Summary}
\vskip 0.3 cm

We have discussed the spin effects in the $\tau$ pair production at LHC.
A few distributions presented here, sensitive to the
$\tau^+\tau^-$ spin correlations,
can be possibly  used for the MSSM Higgs boson searches scenarios at LHC
 to enhance 
sensitivity of the signal or to verify the hypothesis
of the spin zero nature of the Higgs boson.

We have extended the algorithm for the interfacing the $\tau$ lepton decay package 
{\tt TAUOLA} with ``any'' production generator to  include effects due to spin
in elementary $Z/\gamma^{*} \to \tau^+\tau^-$ process.
The interface is  based exclusively on the information stored in the {\tt HEPEVT} 
common block. This code is publicly available {\it e.g.} from the URL address 
\cite{TauolaInterface}.

\vspace{4mm}
\noindent {\bf \large Acknowledgements}\\

Z. W. acknowledges support of the Z\"u{}rich ETH group at CERN while part of this 
work was done.
\\ 
\begin{center}
{ \Large \bf Appendix }
\end{center}
\vskip 0.5 cm
\begin{center}
{\large \it Universal interface for {\tt TAUOLA} package }
\end{center}
\vskip 0.3 cm

Let us recall first some details of 
the universal interface for {\tt TAUOLA} and ``any'' $\tau$ 
production generator
as described in  Ref.~\cite{Golonka:2000iu}. 
The interface uses as an input, the {\tt HEPEVT} common block and
 operates on its content
only. As a demonstration example the
interface is combined  with the {\tt JETSET} generator, however it should work
in the same manner with the {\tt PYTHIA\footnote{It was already checked to be the case.}, 
HERWIG} or {\tt ISAJET} generators as well.

The interface acts in the following way:
\begin{itemize}
\item
The $\tau$ lepton should be forced to be stable in the
 package performing generation 
of the $\tau$ production.
\item
The content of the {\tt HEPEVT} common block is searched 
for all $\tau$ leptons and $\tau$ neutrinos. 
\item
It is checked if there are $\tau$ flavour pairs (two $\tau$ leptons or
$\tau$ lepton and $\tau$ neutrino) originating from the same mother. 
\item
The decays of the $\tau$ flavour pairs are performed with the subroutine {\tt TAUOLA}.
Longitudinal spin correlations are generated  in the case of the $\tau$
 produced from decay of:
$W \to \tau \nu$, $Z/\gamma \to \tau \tau$, 
the neutral Higgs boson $H \to \tau \tau$, and the charged Higgs boson
 $H^{\pm}\to  \tau \nu$.
Parallel or anti-parallel spin configurations are
generated, before calling on the $\tau$ decay, and then the decays
 of 100 \% polarised $\tau$'s are executed. 
\item
In the case of the Higgs boson (for the spin correlations to be generated)
the identifier of the $\tau$ mother must be that of the Higgs boson.
Here, the particle code convention as that used by the {\tt PYTHIA 5.7} 
Monte Carlo is adopted. 
\item
In case of the $W$ and $Z/\gamma$ it is not necessary. 
If from the same mother as that of the $\tau$, a $\nu_\tau$ is also  produced, 
the $W$ is assumed as the mother of the  $\tau$. 
Similarly, if from the same mother another $\tau$ with opposite charge is produced,
the  $Z/\gamma$ is assumed to be the mother of the  $\tau$ pair.
\item
Photon radiation in the decay is performed with {\tt PHOTOS} package 
\cite{Barberio:1990ms,photos:1994}.
\item
Let us note that the calculation of the $\tau$ polarisation created from
the $Z$ and/or virtual $\gamma$ (as function of the direction) represents 
a rather non-trivial extension.
Generally, the dedicated study of 
the production matrix elements of the host generator is necessary in every individual
case. 
\end{itemize}

For some of the technical details on how to use the interface 
we address the reader to Ref. \cite{Golonka:2000iu}. For general 
documentation of {\tt TAUOLA} to Ref. 
\cite{tauola:1990,tauola:1992,tauola:1993}. For the use of {\tt PYTHIA}
reference to  \cite{Sjostrand:2000wi} and references therein will be the best. 

Here, let us concentrate on practical details. The {\tt TAUOLA} interface 
is organised in a modular form to be used conveniently in 
``any environments''.

\vskip 0.4 cm
\centerline{\bf initialisation}
\vskip 0.4 cm

Initialisation is performed with the  {\tt CALL TAUOLA(MODE,KEYSPIN)},
 {\tt MODE=-1}. 
All necessary input is directly coded in subroutine {\tt TAUOLA} placed 
in a file {\tt tauface-jetset.f}.

The following input parameters are set at this call (we omit those which are
standard input for {\tt TAUOLA} as defined in its documentation). 
They are hard-coded in the subroutine.   

\def\sstrut{$\strut\atop\strut$}
\vskip 0.1 cm
\vbox{
$$\halign{
\vrule #
   & \hskip5pt  \sstrut  {\tt #} \hfil \vrule
   & \hskip5pt  \vtop{\hsize=11.8cm {\noindent \strut # \strut}}
   & # \vrule\cr
\noalign{\hrule}
& Parameter & Meaning          &\cr
\noalign{\hrule}
&POL     & Internal switch for spin effects in $\tau$ decy. Normally the user
should set {\tt POL=1.0}, and when {\tt POL=0.0} spin polarisation
effects in the decays are absent &\cr
&KFHIGGS(3)&({\tt KF}=25, 35, 36) Flavour code for $h$, $H$ and $A$  &\cr
&KFHIGCH   &({\tt KF}=37) Flavour code for $H^{+}$ &\cr
& KFZ0     &({\tt KF}=23) Flavour code for $Z^{0}$ &\cr
& KFGAM    &({\tt KF}=22) Flavour code for $\gamma$ &\cr
& KFTAU    &({\tt KF}=15) Flavour code for $\tau^{-}$ &\cr
& KFNUE    &({\tt KF}=16) Flavour code for $\nu_{\tau}$ &\cr
\noalign{\hrule}
}$$}

\vskip 0.4 cm
 \centerline{\bf Event generation}
\vskip 0.4 cm

For every event generated by the production generator, 
all $\tau$ leptons will  
be decayed with the single
{\tt CALL TAUOLA(0,KEYSPIN)}, 
({\tt KEYSPIN=1/0} denotes spin effects switched on/off). Then, all $\tau$
leptons, will be first localised, their positions
stored in internal common block {\tt TAUPOS}, and the information
 necessary for calculation 
of  $\tau$ spin state will be read in, from {\tt HEPEVT} common block. 
Later spin state for the given  $\tau$  (or  $\tau$ pair) 
will be generated, and finally decay of polarised $\tau$ will be  
performed with the standard 
{\tt TAUOLA} action. In 
particular, the decay products of $\tau$ will be boosted to the laboratory frame 
and added to the complete event configuration stored  in {\tt HEPEVT} common block. 

\vskip 0.4 cm
\centerline{\bf Calculation of the $\tau$ spin state}
\vskip 0.4 cm

Once {\tt CALL TAUOLA(0,KEYSPIN)} is executed and $\tau$ leptons found, 
the spin states need to be calculated.

First,
we look  in {\tt HEPEVT} for the position of  $\tau$'s mothers,  and store them, in matrix
{\tt IMOTHER(20)}. Each mother giving
$\tau$ lepton(s) is stored  only once, independently of the number of produced $\tau$'s.
Later, for every {\tt IMOTHER(i)} we execute the following steps:

\begin{enumerate}
\item
 The daughters which are either $\tau$ leptons, or $\nu_\tau$ are searched for.
\item
 Daughters are combined in pairs, case of more than 1 pair is not expected 
 to be important  and ad hoc pairing is then performed.
\item
 The two main cases are thus ($\tau $  $\tau $)  or ($\tau$ $\nu $).
\item
 The default choices are, respectively, $Z/\gamma$ or $W$, unless 
the identifier of
  {\tt IMOTHER(i)} is explicitely that of neutral (or charged) Higgs boson.
\item 
Calculation of the spin parameters is kinematics independent and straightforward
in all cases except  $Z/\gamma$ (see Section 2 for details on physics).
\item
For $Z/\gamma$, the $P_{Z}$ is calculated with the help of the function 
{\tt PLZAPX(HOPE,IM0,NP1,NP2)}. The {\tt HOPE} is the logical parameter defined in subroutine {\tt TAUOLA} placed in file {\tt tauface-} \\{\tt jetset.f}.
It tells  whether spin effects can be calculated or not. It is set to {\tt .false.}, 
if available information is incomplete, then, 
 {\tt PLZAPX(HOPE,IM0,NP1,NP2)} returns $0.5$. The  {\tt IM0} denotes position of the 
$\tau$ mother in {\tt HEPEVT}
common block {\tt NP1} position of  $\tau^{+}$ and  {\tt NP2} of  $\tau^-$.
\item
  To calculate reduced $2 \to 2 $ body kinematical variables $s$ and 
$\cos\theta$ 
subroutine {\tt ANGULU(PD1,PD2,Q1,Q2,COSTHE)} is used. 
4-momenta of the incoming effective beams and outgoing 
$\tau^+$ and $\tau^-$ are denoted by
$\tt PD1, PD2, Q1, Q2$, respectively.
\end{enumerate}

\vskip 0.4 cm
\centerline{\bf Run summary}
\vskip 0.4 cm

After the series of events is generated the optional
  {\tt CALL TAUOLA(1,KEYSPIN)} can be executed.
The  information on the whole sample, such as number of the 
generated $\tau$ decays,
branching ratios calculated from matrix elements {\it etc.}, 
 will be printed.

\vskip 0.4 cm
\centerline{\bf Demonstration program}
\vskip 0.4 cm

Our main program  {\tt demo.f} is stored in subdirectory {\tt demo-jetset}. It reads in
the file {\tt init.dat} which includes some  input parameters for the particular run,
such as number of events to be generated (by {\tt JETSET/PYTHIA}), or the 
type of the interaction it should use to produce $\tau$'s,
 {\it etc}. We address the reader directly to the
code for more details. It is self-explanatory.


\end{document}